\documentclass[twocolumn,tighten]{aastex62}
\usepackage{amssymb}
\usepackage{amsmath}
\usepackage{graphicx}
\usepackage{hyperref}
\usepackage[T1]{fontenc}
\definecolor{revcol}{RGB}{216, 18, 125}
\def\revised#1{#1}
\newcommand{\removed}[1]{}
\newcommand{\todo}[1]{}
\newcommand{\todonext}[1]{}
\def\aap{A\&A}

\def\apjl{ApJL}
\def\apjs{ApJS}

\def\fwhm{FWHM}
\def\kapabs{2.0}
\def\xidust{1.67}
\def\optthin{\mathrm{thin}}

\definecolor{intrevcolor}{rgb}{0.0, 0.0, 1}
\definecolor{newstuffcolor}{rgb}{0.0, 0.6, 0}
\newcommand{\afterintrev}[1]{#1}
\newcommand{\newstuff}[1]{#1}
\def\colsrcname{(1)}
\def\colring{(2)}
\def\colringname{(3)}
\def\colbeam{(4)}
\def\coldomain{(5)}
\def\cola{(6)}
\def\coladec{(7)}
\def\colr0{(8)}
\def\colsigwidth{(9)}
\def\colwd{(10)}
\def\coltd{(11)}
\def\colbnu{(12)}
\def\colwdhp{(13)}
\def\colssb{(14)}
\def\coltaupk{(15)}
\def\colmdthin{(16)}
\def\colmdtrue{(17)}
\def\colwmin{(4)}
\def\colwmax{(5)}
\def\colsiggmin{(6)}
\def\colsiggmax{(7)}
\def\colamax{(8)}
\def\colstexmp{(9)}
\def\colwdw{(10)}
\def\colastwmax{(11)}
\def\colastwmin{(12)}
\def\colaexmp{(13)}
\begin{document}

\title{The Disk Substructures at High Angular Resolution Project (DSHARP)\\ VI: Dust trapping in thin-ringed protoplanetary disks}
\shorttitle{Dust trapping}
\shortauthors{Dullemond et al.}

\correspondingauthor{Cornelis Dullemond}
\email{dullemond@uni-heidelberg.de}

\author[0000-0002-7078-5910]{Cornelis P.~Dullemond}
\affiliation{Zentrum f\"ur Astronomie, Heidelberg University, Albert Ueberle Str.~2, 69120 Heidelberg, Germany}

\author[0000-0002-1899-8783]{Tilman Birnstiel}
\affiliation{University Observatory, Faculty of Physics, Ludwig-Maximilians-Universit\"at M\"unchen, Scheinerstr.~1, 81679 Munich, Germany}

\author[0000-0001-6947-6072]{Jane Huang}
\affiliation{Harvard-Smithsonian Center for Astrophysics, 60 Garden Street, Cambridge, MA 02138, USA}

\author{Nicol\'as T.~Kurtovic}
\affiliation{Departamento de Astronom\'ia, Universidad de Chile, Camino El Observatorio 1515, Las Condes, Santiago, Chile}

\author{Sean M.~Andrews}
\affiliation{Harvard-Smithsonian Center for Astrophysics, 60 Garden Street, Cambridge, MA 02138, USA}

\author[0000-0003-4784-3040]{Viviana V.~Guzm\'an}
\affiliation{Joint ALMA Observatory, Avenida Alonso de C\'ordova 3107, Vitacura, Santiago, Chile; Instituto de Astrof\'isica, Pontificia Universidad Cat\'olica de Chile, Av. Vicu\~na Mackenna 4860, 7820436 Macul, Santiago, Chile}

\author[0000-0002-1199-9564]{Laura M. P\'erez}
\affiliation{Departamento de Astronom\'ia, Universidad de Chile, Camino El Observatorio 1515, Las Condes, Santiago, Chile}

\author[0000-0001-8061-2207]{Andrea Isella}
\affiliation{Department of Physics and Astronomy, Rice University 6100 Main Street, MS-108, Houston, TX 77005, USA}

\author{Zhaohuan Zhu}
\affiliation{Department of Physics and Astronomy, University of Nevada, Las Vegas, 4505 S. Maryland Pkwy, Las Vegas, NV, 89154, USA}

\author[0000-0002-7695-7605]{Myriam Benisty}
\affiliation{Unidad Mixta Internacional Franco-Chilena de Astronom\'{i}a, CNRS/INSU UMI 3386, Departamento de Astronom\'ia, Universidad de Chile, Camino El Observatorio 1515, Las Condes, Santiago, Chile; Univ. Grenoble Alpes, CNRS, IPAG, 38000 Grenoble, France}

\author[0000-0003-1526-7587]{David J.~Wilner}
\affiliation{Harvard-Smithsonian Center for Astrophysics, 60 Garden Street, Cambridge, MA 02138, USA}

\author[0000-0003-1172-3039]{Xue-Ning Bai}
\affiliation{Institute for Advanced Study and Tsinghua Center for Astrophysics, Tsinghua University, Beijing 100084, China}

\author[0000-0003-2251-0602]{John M.~Carpenter}
\affiliation{Joint ALMA Observatory, Avenida Alonso de C\'ordova 3107, Vitacura, Santiago, Chile}

\author{Shangjia Zhang}
\affiliation{Department of Physics and Astronomy, University of Nevada, Las Vegas, 4505 S. Maryland Pkwy, Las Vegas, NV, 89154, USA}

\author{Luca Ricci}
\affiliation{Department of Physics and Astronomy, California State University Northridge, 18111 Nordhoff Street, Northridge, CA 91130, USA}

\begin{abstract}
A large fraction of the protoplanetary disks observed with ALMA display multiple
well-defined and nearly perfectly circular rings in the continuum, in many cases
with substantial peak-to-valley contrast. The DSHARP campaign shows that several
of these rings are very narrow in radial extent. In this paper we test the
hypothesis that these dust rings are caused by dust trapping in radial pressure
bumps, and if confirmed, put constraints on the physics of the dust trapping
mechanism. We model this process analytically in 1D, assuming axisymmetry. By
comparing this model to the data, we find that all rings are consistent with
dust trapping. Based on a plausible model of the dust temperature we find that
several rings are narrower than the pressure scale height, providing strong
evidence for dust trapping. The rings have peak absorption optical depth in the
range between 0.2 and \revised{0.5}. The dust masses stored in each of these rings is of
the order of tens of Earth masses, though much ambiguity remains due to the
uncertainty of the dust opacities. The dust rings are dense enough to
potentially trigger the streaming instability, but our analysis cannot give
proof of this mechanism actually operating. Our results show, however, that the
combination of very low $\alpha_{\mathrm{turb}}\ll 5\times 10^{-4}$ and very
large grains $a_{\mathrm{grain}}\gg 0.1\,\mathrm{cm}$ can be excluded by the
data for all the rings studied in this paper.
\end{abstract}

\keywords{}

\section{Introduction}
The concept of dust trapping in local pressure maxima has become a central theme
in studies of planet formation and protoplanetary disk evolution, because it might
provide an elegant solution to several problems in these fields of
study. Theories of planet formation are plagued by the ``radial drift barrier'':
the problem that, as dust aggregates grow by coagulation, they tend to radially
drift toward the star before they reach planetesimal size
\citep[e.g.][]{2010A&A...513A..79B}. A natural solution to this problem could be
the trapping of dust particles in local pressure maxima
\citep{1972fpp..conf..211W, 2007ApJ...664L..55K, 1995A&A...295L...1B,
  1997Icar..128..213K}. Not only does this process prevent excessive radial
drift of dust particles, it also tends to concentrate the dust into small
volumes and high dust-to-gas ratios, which is beneficial to planet formation.
From an observational perspective, the radial drift
problem manifests itself by the presence of large grains in the outer regions of
protoplanetary disks \citep{2003A&A...403..323T, 2009ApJ...700.1502A,
2010A&A...512A..15R}, which
appears to be in conflict with theoretical predictions
\citep{2007A&A...469.1169B}. One possible solution to this observational
conundrum could be that the disks are much more massive in the gas than
previously suspected, leading to a higher gas friction for millimeter grains and
thus longer drift time scales \citep{2017ApJ...840...93P}.

Another explanation is to invoke dust traps. The most striking observational
evidence for dust trapping seems to come from large transitional disks, which
feature giant dust rings, sometimes lopsided, in which large quantities of dust
appears to be concentrated \citep{2013Natur.493..191C,
  2013Sci...340.1199V}. These observations appear to be well explained by the
dust trapping scenario \citep{2012A&A...545A..81P}. But these transitional disks
seem to be rather violent environments, possibly with strong warps
\citep{2015ApJ...798L..44M, 2017A&A...597A..42B} and companion-induced spirals
\citep{2016ApJ...816L..12D}.

For more ``normal'' protoplanetary disk the dust traps would have to be more
subtle. \citet{2012A&A...538A.114P} explored the possibility that the disk
contains many axisymmetric local pressure maxima, and calculated how the dust
drift and growth would behave under such conditions. It was found that, if the
pressure bumps are strong enough, the dust trapping can keep a sufficient
fraction of the dust mass at large distances from the star to explain the
observed dust millimeter flux. It would leave, however, a detectable pattern of
rings that should be discernable with ALMA observations.  Since the multi-ringed
disk observation of HL Tau \citep{2015ApJ...808L...3A} a number of such
multi-ringed disks have been detected \citep{2016ApJ...820L..40A,
  2016PhRvL.117y1101I, 2017ApJ...851L..23C, 2017A&A...600A..72F,
  2018A&A...610A..24F, 2018MNRAS.475.5296D, 2018A&A...616A..88V,
  2018ApJ...866L...6C, 2018arXiv181006044L}. It is therefore very tempting to see also these
multi-ringed disks as evidence for dust trapping, and as an explanation for the
retention of dust in the outer regions of protoplanetary disks.

The data from the ALMA Large Programme DSHARP \citep{dsharp:andrews} offers an
exciting new opportunity to put this concept to the test, and to put constraints
on the physics of dust trapping in axisymmetric pressure maxima. This is an
opportunity which we explore in this paper.

As is shown by \citet{dsharp:huangrings}, most of the disks in the DSHARP sample
display multi-ringed substructure. We investigate whether these rings are caused
by dust trapping, and if so, what we can learn about dust trapping from these
data. We will focus on a subsample of rings, for which the contrast is
particularly strong, so that amplitude and width can be clearly defined.  We
study the rings individually, assuming that the dust does not escape from the
ring. This makes it possible to look for a steady-state dust trapping solution
in which the radial drift forces (that push the dust to the pressure peak) are
balanced by turbulent mixing (that tends to smear out the dust away from the
pressure peak). In Appendix \ref{sec-steady-state-analytic-trap-model} we will
construct a very simplified analytic dust trapping model, and confront this with
the most well-isolated rings from our sample.

\todo{Check the new structure of the paper.}

The structure of the paper is as follows. We first review, in Section
\ref{sec-data}, our subsample of rings, and how the radial profile of the
intensity was obtained. Next we fit these rings to Gaussians (Section
\ref{sec-gauss-fits}), because this will make the quantitative analysis of the
subsequent sections easier. In Section \ref{sec-opt-thin-analysis} we will first
analyze these Gaussian fits under the assumption that these rings are optically
thin. It turns out, however, that the optical depths are on the border between
thin and thick, requiring us to explore, in Section
\ref{sec-optical-depth-effects} how moderate optical depths affect our results,
and correct for this. We are then ready to compare this to a model of dust
trapping. In Section \ref{sec-rings-as-dust-traps} we take the simplest possible
model of dust trapping: that of a Gaussian pressure bump. This allows us to
derive most results analytically. In Section \ref{sec-planet-gap} we go one step
further by numerically exploring dust trapping by a very simple planetary gap
model, and see to which extent the results are different and may fit better or
worse to the data. We close with a discussion and conclusion section.

\section{The high-contrast rings of AS 209, Elias 24, HD 163296, GW Lup and HD 143006}
\label{sec-data}
In this paper we focus on a subsample of sources of the DSHARP Programme
that show high-contrast and radially thin rings that are separated by deep
valleys, and that are sufficiently face-on to not have to worry much about 3-D
line-of-sight issues. These are AS 209, Elias 24, HD 163296, GW Lup and
HD 143006. Their stellar parameters are given in Table \ref{tab-stellar-params}.

\begin{deluxetable}{cccccc}
\tablecaption{The stellar parameters assumed for the stars
  studied in this paper, and the ALMA beam size and position angle of the DSHARP
  observations.\label{tab-stellar-params}}
\tablecolumns{6}
\tablewidth{0pt}
\tablehead{
\colhead{Source} &
\colhead{$d$} &
\colhead{$M_{*}$} &
\colhead{$L_{*}$} &
\colhead{$i$} &
\colhead{Beam, PA} \\
\colhead{} &
\colhead{[pc]} &
\colhead{$[M_{\odot}]$} &
\colhead{$[L_{\odot}]$} &
\colhead{[deg]} &
\colhead{[mas], [deg]} \\
}
\startdata
AS 209     & 121              & 0.83        &  1.41   & 35 &  38$\times$36, 68 \\ 
Elias 24   & 136              & 0.78        &  6.0    & 29 & 37$\times$24, 82 \\ 
HD 163296  & 101              & 2.04         &  17.0  & 47 &  48$\times$38, 82 \\ 
GW Lup     & 155              & 0.46        &  0.33   & 39 &  45$\times$43, 1  \\ 
HD 143006  & 165              & 1.78        &  3.80   & 19 &  46$\times$45, 51 \\ 
\enddata
\tablecomments{Distance is in parsec and mass and luminosity are in units of
  the solar values. The beam is in milliarcsecond. Inclination and position angle
  are in degrees (PA east from north for the major axis). More details, as well as references and
  uncertainty estimates, can be found in \citet{dsharp:andrews}.}
\end{deluxetable}

\begin{figure*}
\centerline{\includegraphics[width=0.95\textwidth]{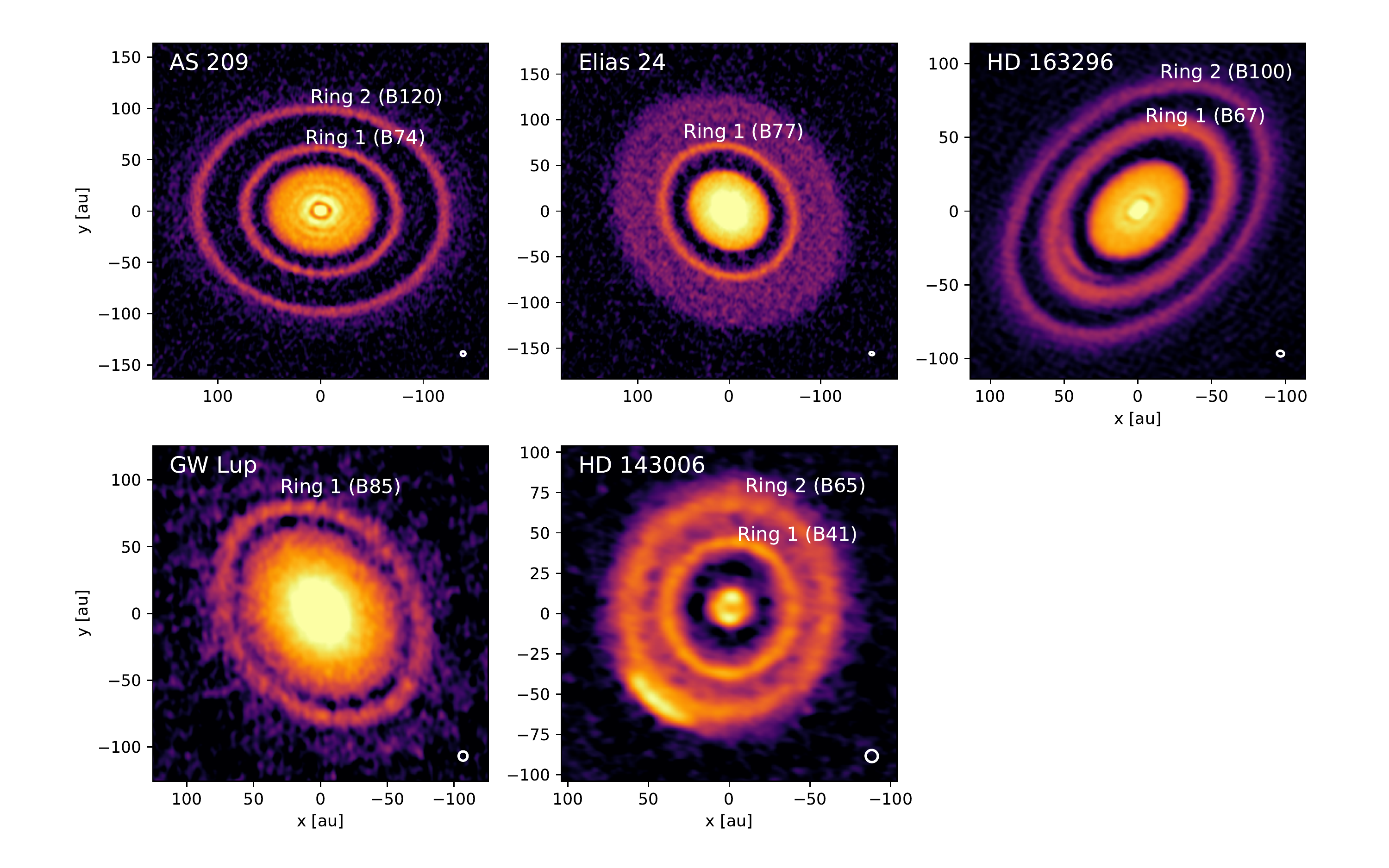}}
\caption{\label{fig-obs-images}The continuum maps in band 6 of the five disks in
  our sample which have the most pronounced rings.  The eight highest contrast
  rings, which are the topic of this paper, are marked in the images. The
  color scale is the same as from \citet{dsharp:huangrings}. For a
  detailed description of these data, see \citet{dsharp:guzman} for AS 209,
  \citet{dsharp:isella} for HD 163296, 
  \afterintrev{\citet{dsharp:perez} for HD 143006,
    and \citet{dsharp:huangrings} for the rest.}}
\end{figure*}

A gallery of these sources is shown in Fig.~\ref{fig-obs-images}.  For an
overview of the ALMA Large Programme we refer to \citet{dsharp:andrews}, and for
an in-depth discussion on the data of the individual sources we refer to
\citet{dsharp:huangrings}, \citet{dsharp:isella}, \citet{dsharp:guzman}
\afterintrev{and \citet{dsharp:perez}}.

\begin{figure*}
\centerline{\includegraphics[width=0.85\textwidth]{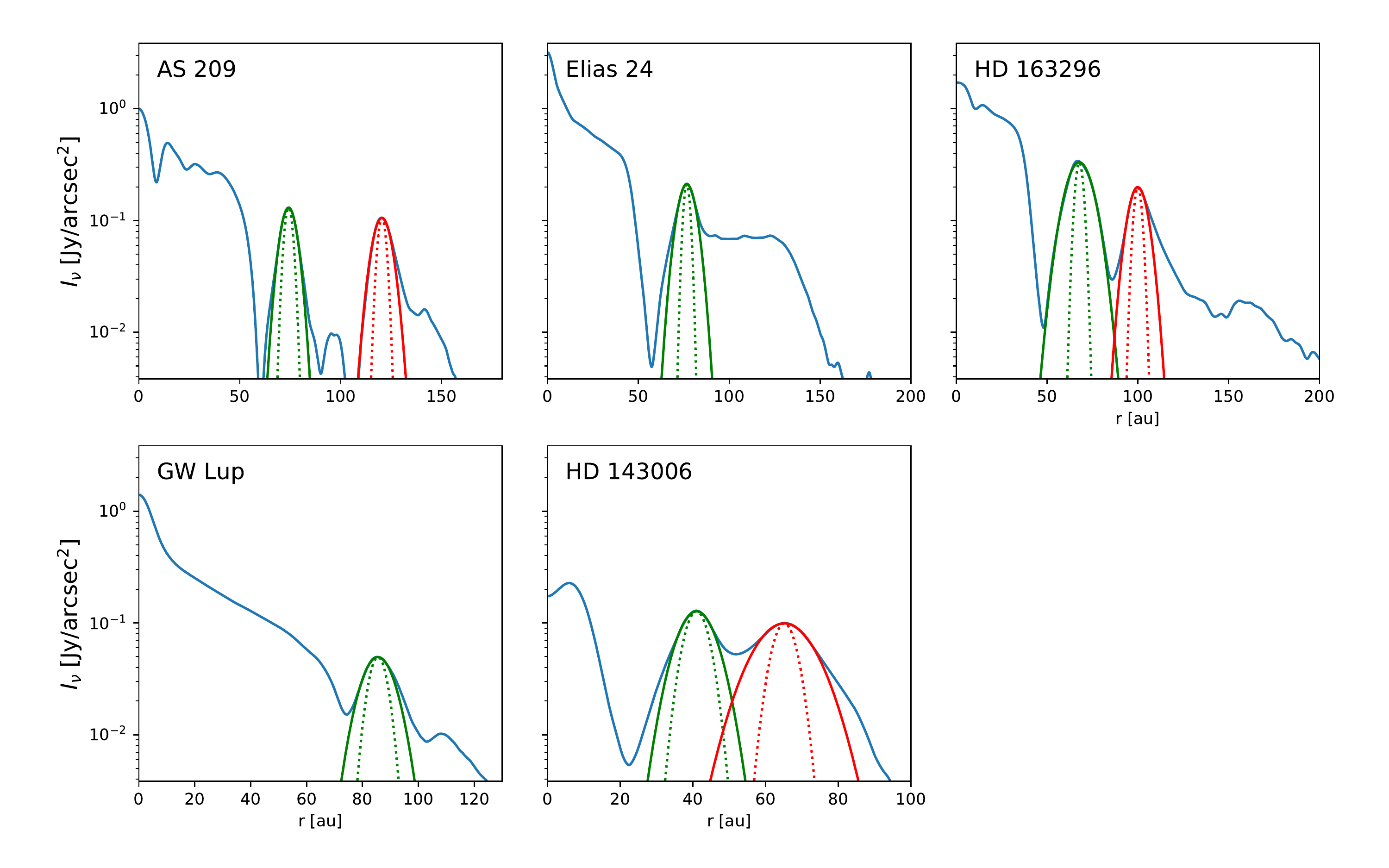}}
\caption{\label{fig-obs-profiles}The \afterintrev{intensity profiles} in
  band 6 of the five disks in our sample which have the most pronounced rings.
  The vertical axis is logarithmic to better show the contrast.
  The eight highest contrast rings are fitted by a Gaussian profile, shown
  as the solid inverse parabolas. The dotted inverse parabolas are Gaussians
  with the width of the ALMA beam. For a detailed description of these
  data, see \citet{dsharp:huangrings}. \afterintrev{The unit of intensity
    is always $\mathrm{Jy/arcsec}^2$ at $\lambda=0.125\,\mathrm{cm}$.
    For Elias 24 the observations \afterintrev{had a central wavelength of}
    $\lambda=0.129\,\mathrm{cm}$,
    but we rescaled to $\lambda=0.125\,\mathrm{cm}$ assuming a spectral
  slope of $I_\nu\propto \nu^2$, meaning a 6.5\% increase.}}
\end{figure*}

The high-contrast rings of these sources provide ``clean laboratories'' for
testing the theory of dust trapping in a ring-by-ring manner.
Fig.\ \ref{fig-obs-profiles} shows the radial profile (deprojected for
inclination) of the thermal emission of the dust of the five
disks. \afterintrev{These brightness profiles are expressed as intensity $I_\nu$
  in units of $\mathrm{Jy}/\mathrm{arcsec}^2$.} The procedure used to extract
these radial profiles from the continuum maps is described by
\citet{dsharp:huangrings}. \afterintrev{In creating these profiles, the
  ``arcs'' seen in HD 163296 and HD 143006 were excised, so these radial
profiles represent the axially symmetric structures only.}

The DSHARP sample has many more sources with rings, and several of the sources
we study in this paper display more than just the one or two rings we focus on
\citep{dsharp:huangrings}. Particularly striking in this regard is AS 209, which
features three more ringlike structures in the inner disk. The contrast and
radial separation of these rings is, however, much less than for the subset of
rings we choose for this paper. While dust trapping can certainly also play a
role in those rings, it is much harder to quantify this. For that reason we do
not consider those rings further in this paper.

\section{Fitting a Gaussian profile to the ring emission}
\label{sec-gauss-fits}
As we will discuss later (Section \ref{sec-rings-as-dust-traps}), for a radially
Gaussian pressure bump the solution to the radial dust mixing and drift problem
is, to first approximation, also a Gaussian surface density profile. It has a
width smaller than, or equal to, that of the gas pressure bump. Our analysis of
the eight rings of this paper therefore naturally starts with the fitting of the
observed radial intensity profiles with a Gaussian function. We choose here to
do so in the image plane, because that allows us to select an individual ring,
and study it independently of the emission elsewhere. But note that other papers
in the DSHARP series have done, for individual sources, fits in the uv plane
\citep{dsharp:guzman,dsharp:isella,dsharp:perez}.

\subsection{\revised{Procedure}}
\begin{figure*}
\centerline{\includegraphics[width=0.85\textwidth]{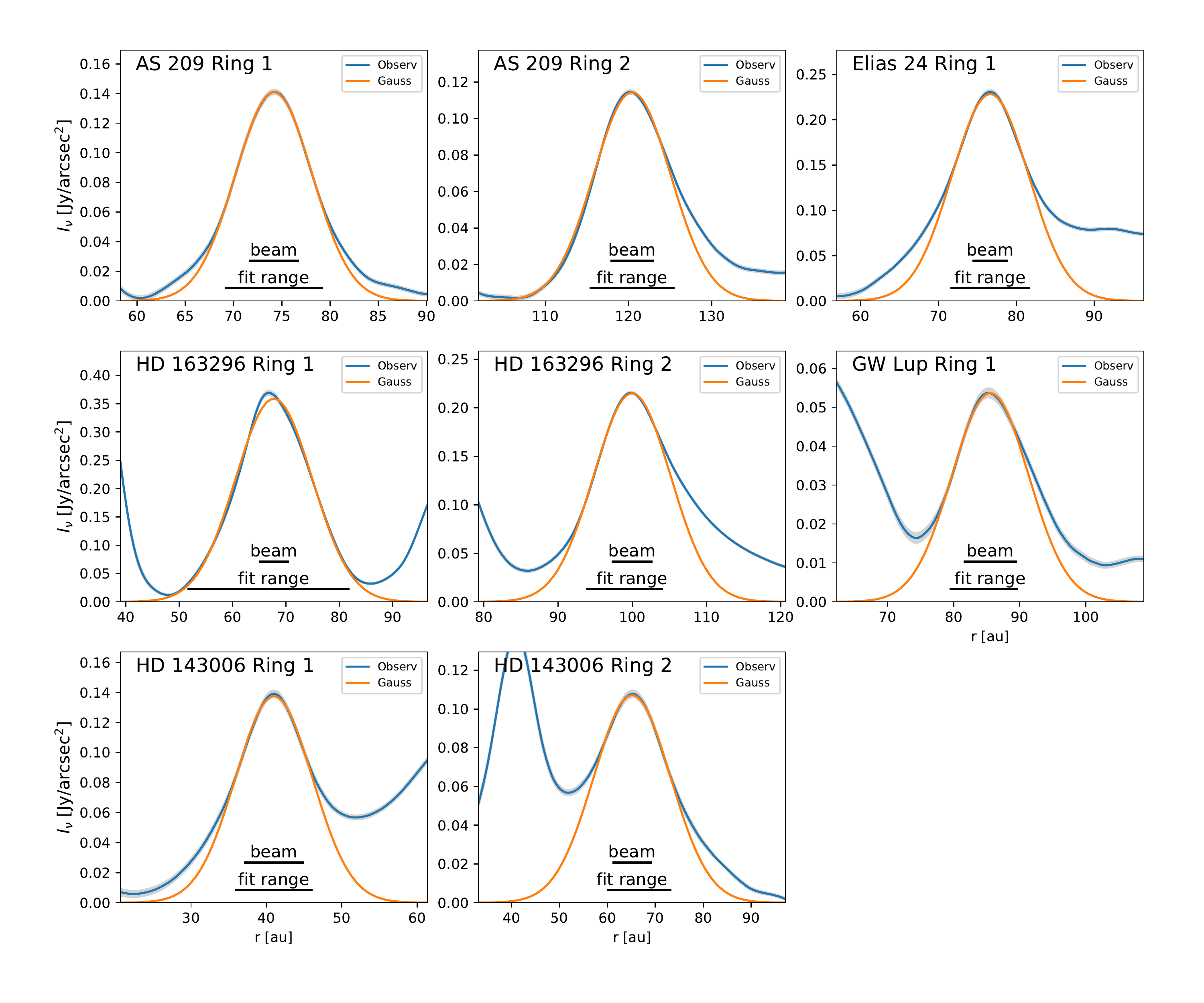}}
\caption{\label{fig-obs-gaussfits}Gaussian fits to the eight rings of this
  paper. The blue curves are the observations, the orange curves are the best fit
  Gaussian profiles. The ``fit range'' bar shows the radial range within which
  the Gauss \afterintrev{curve} was fitted to the data. The fit range was chosen to fit the part of
  the curve that, by eye, most resembles a Gaussian. The ``beam'' bar shows the
  \fwhm{} beam size of the observations. The grey band around the blue curve shows
  the estimated uncertainty of the data.}
\end{figure*}

The aim is to find, for each ring, a Gaussian intensity
profile
\begin{equation}\label{eq-gauss-lin-br-temp}
I_{\nu}^{\mathrm{gauss}}(r) = A\,\exp\left(-\frac{(r-r_0)^2}{2\sigma^2}\right)
\end{equation}
that best describes the ring. To be more precise: We determine the values of
$A$, $r_0$ and $\sigma$ for which Eq.~(\ref{eq-gauss-lin-br-temp}) best fits the
observed \afterintrev{intensity} profile $I_{\nu}^{\mathrm{obs}}(r)$ shown in
Fig.~\ref{fig-obs-profiles} within a prescribed radial domain as given in Table
\ref{tab-gauss-params}. Details of the fitting procedure are described in
Appendix \ref{sec-gauss-fitting-procedure} and \ref{sec-comments-on-gauss}. The
Gaussian fits appear as inverse parabolas in Fig.\ \ref{fig-obs-profiles}.
In \revised{the close-up views of Fig.~\ref{fig-obs-gaussfits} they are
  overplotted in orange.} The parameters of the best fits are listed in
\revised{Tables \ref{tab-gauss-params} and \ref{tab-gauss-fit-errors}.}

\begin{figure*}
  \centerline{
    \includegraphics[width=0.42\textwidth]{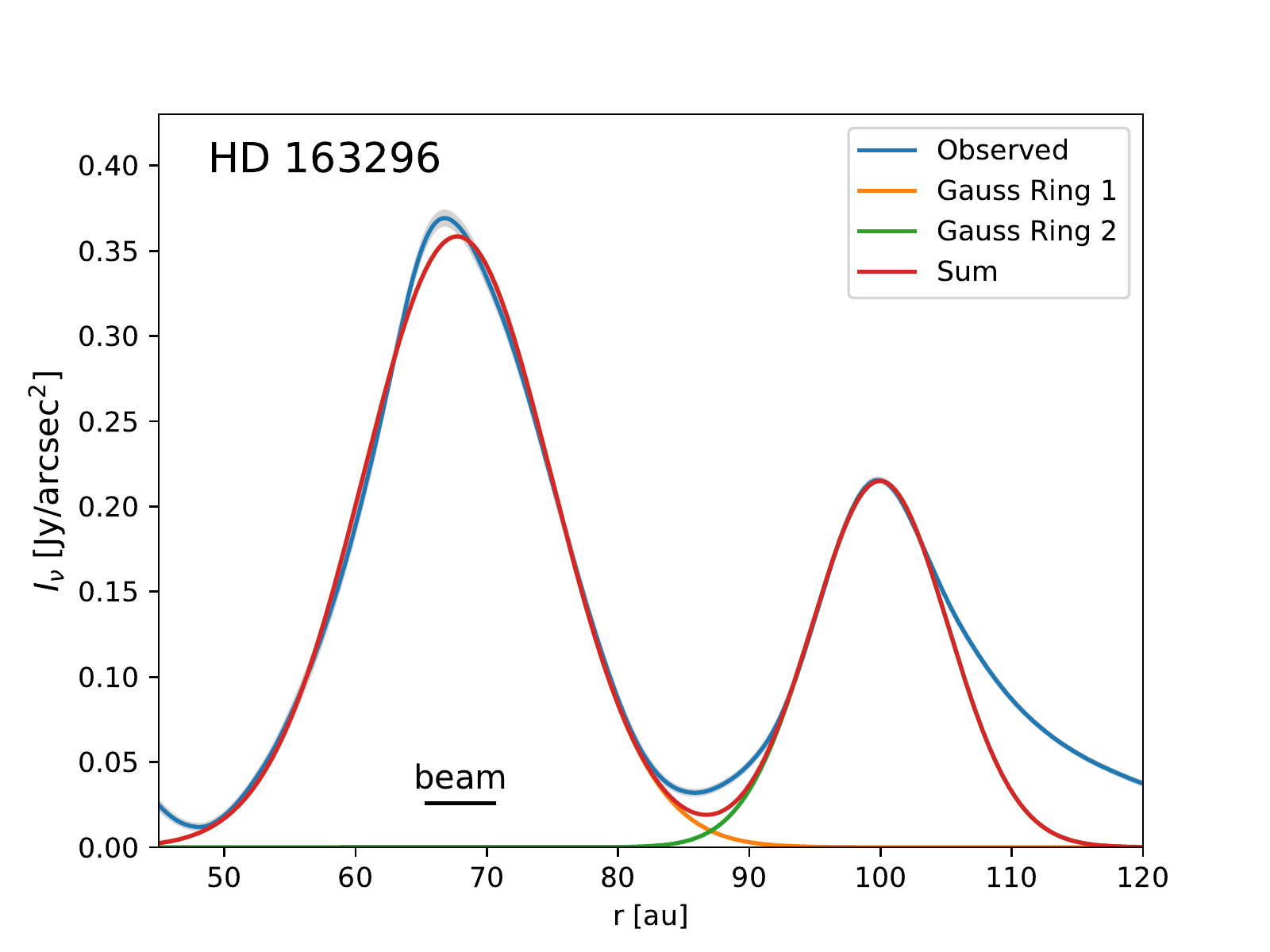}
    \includegraphics[width=0.42\textwidth]{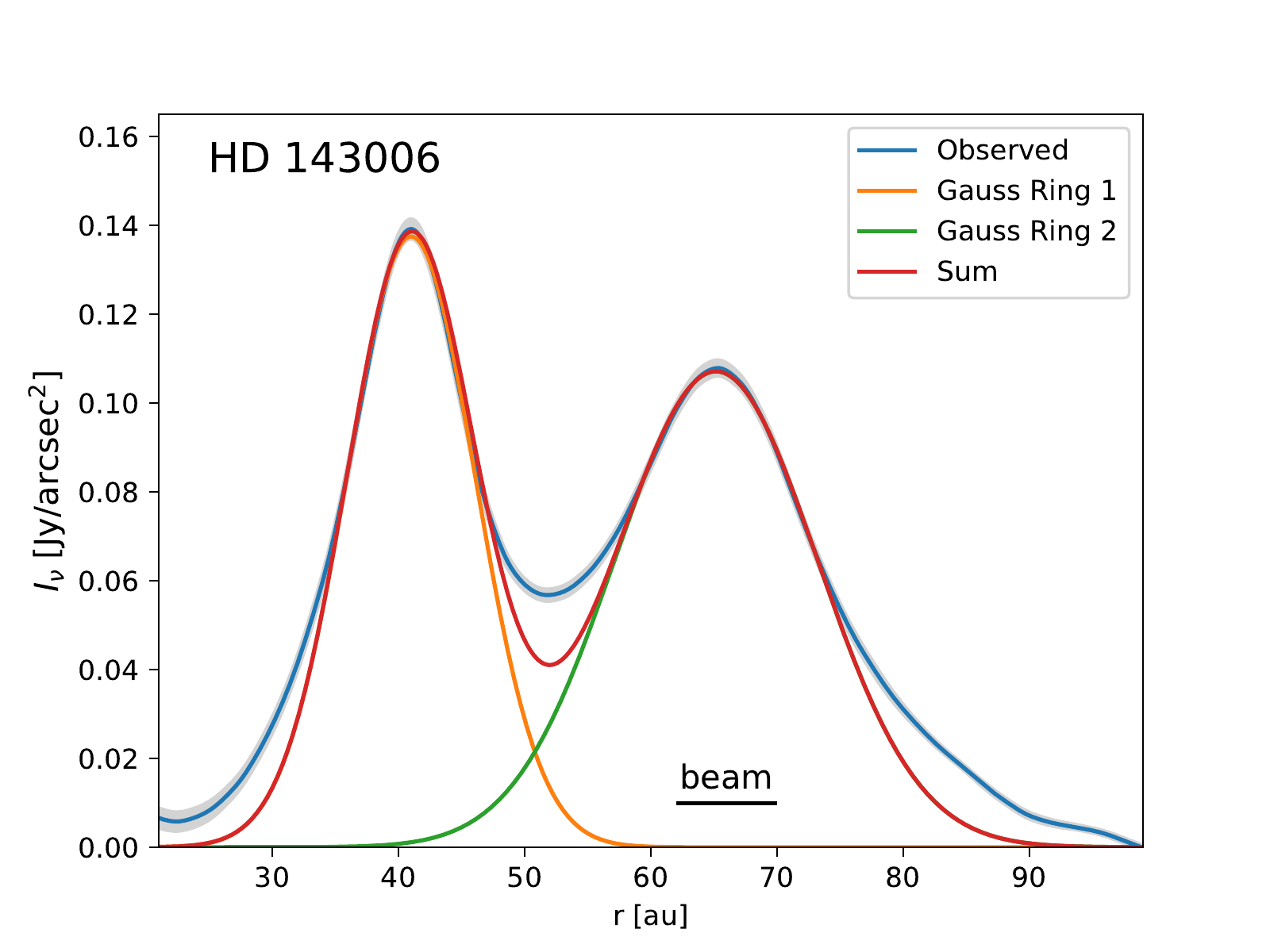}
  }
  \caption{\label{fig-obs-doublegauss}The sum of the two Gaussian fits
    for the two sources with two partly overlapping rings: HD 163296 and
    HD 143006.}
\end{figure*}

\begin{deluxetable*}{ccccccccccccccccc}[b!]
\tablecaption{The model parameters for the Gaussian ring fits
in Figs.~\ref{fig-obs-profiles} and \ref{fig-obs-gaussfits}\label{tab-gauss-params}}
\tablecolumns{17}
\tablewidth{0pt}
\tablehead{
\colhead{Source} &
\colhead{Ring} &
\colhead{Name} &
\colhead{Beam} &
\colhead{Domain} &
\colhead{$A$} &
\colhead{$A_{\mathrm{dec}}$} &
\colhead{$r_0$} &
\colhead{$\sigma$} &
\colhead{$w_d$} &
\colhead{$T_{\mathrm{d}}$} &
\colhead{$B_{\nu}(T_{\mathrm{d}})$} &
\colhead{$w_d/h_p$} &
\colhead{$\sigma/\sigma_b$} &
\colhead{$\tau_\nu^{\mathrm{peak}}$} &
\colhead{$M_{\mathrm{d}}^{\mathrm{thin}}$} &
\colhead{$M_{\mathrm{d}}^{\mathrm{true}}$} \\
\colhead{} &
\colhead{} &
\colhead{} &
\colhead{$[\mathrm{mas}]$} &
\colhead{$[\mathrm{au}]$} &
\colhead{$[\mathrm{Jy/as}^2]$} &
\colhead{$[\mathrm{Jy/as}^2]$} &
\colhead{$[\mathrm{au}]$} &
\colhead{$[\mathrm{au}]$} &
\colhead{$[\mathrm{au}]$} &
\colhead{$\mathrm{[K]}$} &
\colhead{$[\mathrm{Jy/as}^2]$} &
\colhead{} &
\colhead{} &
\colhead{} &
\colhead{$[M_{\oplus}]$} &
\colhead{$[M_{\oplus}]$} \\
\colhead{\colsrcname} &
\colhead{\colring} &
\colhead{\colringname} &
\colhead{\colbeam} &
\colhead{\coldomain} &
\colhead{\cola} &
\colhead{\coladec} &
\colhead{\colr0} &
\colhead{\colsigwidth} &
\colhead{\colwd} &
\colhead{\coltd} &
\colhead{\colbnu} &
\colhead{\colwdhp} &
\colhead{\colssb} &
\colhead{\coltaupk} &
\colhead{\colmdthin} &
\colhead{\colmdtrue}
}
\startdata
AS 209     & 1 & B74  & 40 &  69 --  79 &   0.14 &   0.17 &  74.2 &   3.98 &   3.38 &  15.8 &   0.45 &   0.6 &   1.9 &   0.46 &  27.0 &  31.5\\
AS 209     & 2 & B120 & 40 & 115 -- 125 &   0.11 &   0.13 & 120.4 &   4.62 &   4.11 &  12.4 &   0.32 &   0.4 &   2.2 &   0.52 &  58.7 &  69.8\\
Elias 24   & 1 & B77  & 31 &  72 --  82 &   0.23 &   0.25 &  76.7 &   4.93 &   4.57 &  22.3 &   0.72 &   0.6 &   2.7 &   0.42 &  35.4 &  40.8\\
HD 163296  & 1 & B67  & 51 &  52 --  82 &   0.36 &   0.38 &  67.7 &   7.18 &   6.84 &  30.8 &   1.06 &   1.6 &   3.2 &   0.44 &  48.3 &  56.0\\
HD 163296  & 2 & B100 & 51 &  94 -- 104 &   0.21 &   0.24 & 100.0 &   5.17 &   4.67 &  25.3 &   0.84 &   0.7 &   2.3 &   0.33 &  39.0 &  43.6\\
GW Lup     & 1 & B85  & 49 &  79 --  89 &   0.05 &   0.06 &  85.6 &   5.81 &   4.80 &  10.2 &   0.24 &   0.6 &   1.8 &   0.32 &  33.2 &  37.0\\
HD 143006  & 1 & B41  & 46 &  35 --  45 &   0.14 &   0.18 &  41.0 &   5.09 &   3.90 &  27.2 &   0.92 &   1.9 &   1.6 &   0.22 &   9.2 &   9.9\\
HD 143006  & 2 & B65  & 46 &  59 --  72 &   0.11 &   0.12 &  65.2 &   8.01 &   7.31 &  21.6 &   0.69 &   2.0 &   2.4 &   0.19 &  24.0 &  25.6\\
\enddata

\tablecomments{\afterintrev{\colring{} Internal numbering of the rings in this
    paper.  \colringname{} Ring name from \citet{dsharp:huangrings}. \colbeam{}
    Effective full-width-at-half-max beam size (see Appendix
    \ref{sec-effective-kernel}).  \coldomain{} Radial fitting range. \cola{}
    Peak intensity $A$ of the best-fit Gaussian ring model. \coladec{}
    Deconvolved peak intensity $A_{\mathrm{dec}}$. \colr0{} Ring radius $r_0$ in
    $\mathrm{au}$. \colsigwidth{} Standard deviation width $\sigma$ in units of
    $\mathrm{au}$. \colwd{} Width $w_d$ of the underlying (deconvolved) dust
    emission profile, also expressed as standard deviation in units of
    $\mathrm{au}$. \coltd{} Midplane temperature $T_{\mathrm{d}}$ of the disk
    (we assume gas and dust temperature to be equal) computed from
    Eq.~(\ref{eq-disk-temperature-model}), assuming a flaring angle of
    $\varphi=0.02$. \colbnu{} Planck function at $T_{\mathrm{d}}$ in band
    6. \colwdhp{} Deconvolved dust ring width $w_d$ in units of the disk
    pressure scale height $h_p$ computed from $T_{\mathrm{d}}$. \colssb{} Ratio
    of observed ring width $\sigma$ to standard deviation beam width $\sigma_b$.
    \coltaupk{} estimated optical depth $\tau_\nu^{\mathrm{peak}}$ at the peak
    of the ring, calculated from Eq.~(\ref{eq-tau-estimate}). \colmdthin{} Dust
    mass estimate $M_{\mathrm{d}}^{\mathrm{thin}}$ using optically thin
    approximation. \colmdtrue{} Dust mass estimate
    $M_{\mathrm{d}}^{\mathrm{true}}$ including optical depth correction.  In
    making these mass estimates we use the DSHARP dust opacity model
    \citep{dsharp:birnstiel} for a grain radius of $a=0.1\,\mathrm{cm}$, which
    yields an absorption opacity
    $\kappa_\nu^{\mathrm{abs}}(\lambda=0.125\,\mathrm{cm})=\kapabs{}\,\mathrm{cm}^2/\mathrm{g}$.}}
\end{deluxetable*}

The observed rings are the result of the thermal emission of a dust ring
convolved with the ALMA beam. To obtain the width of the underlying dust ring we
have to deconvolve. Assuming a Gaussian beam and a Gaussian dust ring, we can
use the rule of the convolution of two Gaussians, and obtain the width $w_d$ of
the dust ring
\begin{equation}\label{eq-simple-deconvolve-gauss}
w_d=\sqrt{\sigma^2-\sigma_b^2}
\end{equation}
where $\sigma_b$ is the beam width expressed as standard deviation in units of
$\mathrm{au}$. The effects of the elliptical shape of the beam and the
inclination of the disk are accounted for in the way described in Appendix
\ref{sec-effective-kernel}. The \revised{resulting} values of
$b_{\mathrm{fwhm,as}}$ are listed in Table \ref{tab-gauss-params}, \revised{and
  the corresponding $\sigma_b$ can be computed through
  $\sigma_b=d_{\mathrm{pc}}b_{\mathrm{fwhm,as}}/2.355$, where $d_{\mathrm{pc}}$
  is the distance to the source in units of parsec.}

The slightly narrower deconvolved ring should also have a correspondingly
higher amplitude $A_{\mathrm{dec}}$ given by
\begin{equation}\label{eq-a-deconv}
A_{\mathrm{dec}} = \frac{\sigma}{w_d} \, A
\end{equation}
to conserve luminosity, where we ignore the geometric effects due to the
circular coordinates. The values of $A_{\mathrm{dec}}$ are listed in Table
\ref{tab-gauss-params} as well.

For completeness, let us note that the deconvolved Gaussian model then becomes
\begin{equation}\label{eq-gauss-lin-br-temp-deconv}
I_{\nu}^{\mathrm{gauss,dec}}(r) = A_{\mathrm{dec}}\,\exp\left(-\frac{(r-r_0)^2}{2w_d^2}\right)
\end{equation}

\subsection{\revised{Results}}
\revised{The immediate first result is that we see that all the rings are
  radially resolved by our observations. If the dust rings were much narrower
  than the beam ($w_d\ll \sigma_b$), then this would have been apparent by
  having $\sigma\simeq \sigma_b$. Although the ratio $\sigma/\sigma_b$ (column
  14 in Table \ref{tab-gauss-params}) is in some cases less than 2, it is in all
  cases clearly larger than 1. For this reason 
  Eq.~(\ref{eq-simple-deconvolve-gauss}) produces reasonably reliable values
  for the widths $w_d$ of the underlying dust rings.}

One of the most important pieces of information we can \revised{now} derive from
these Gaussian fits is the ratio of the ring width $w_d$ to the local pressure
scale height $h_p$. If this ratio is substantially less than 1, dust trapping
must be at work, as we will argue below. Unfortunately, $h_p$ can only be
estimated, because we do not know the disk midplane temperature very well. From
the continuum images we have no information about $T_d(r)$. From the $^{12}$CO
line emission one can estimate the temperature in the disk surface layers, but
\revised{it is much more difficult to do that for}
the midplane \citep[see e.g.][]{2018ApJ...853..113W}. We will instead
estimate the midplane disk temperature using the following simple irradiated
flaring disk recipe:
\begin{equation}\label{eq-disk-temperature-model}
T_{\mathrm{d}}(r) = \left(\frac{\tfrac{1}{2}\varphi L_{*}}{4\pi r^2\sigma_{\mathrm{SB}}}\right)^{1/4}
\end{equation}
where $\sigma_{\mathrm{SB}}$ is the Stefan-Boltzmann constant and $\varphi$ is
the so-called flaring angle \citep[e.g.][]{1997ApJ...490..368C,
  1998ApJ...500..411D, 2001ApJ...560..957D}.\removed{The factor of $1/2$ in front of the
flaring angle $\varphi$ originates from the consideration that the stellar
radiation heats the surface layer of the disk, which then radiates half of that
energy away and half into the deeper regions of the disk. Only the latter half
goes into the energy balance equation between irradiation and radiative cooling
of the disk interior.} We take the flaring angle to be $\varphi=0.02$ which is an
estimate based on typical values from models. The resulting values
  of $T_d$ at the peak of the rings are given in Table \ref{tab-gauss-params}.
\removed{In using this simple \afterintrev{flaring angle} recipe we assume that the
optical and infrared dust opacity is dominated by a population of small grains
that is vertically more extended than the larger dust grains seen in the ALMA
images. \afterintrev{There is ample evidence for this from scattered light images
  of such protoplanetary disks, where the vertically extended geometry is
  clearly seen \citep[e.g.][]{2018arXiv180310882A}.}  The two dust populations
may be largely decoupled (see e.g.\ the BL-series of models of
\citet{2004A&A...417..159D} as an illustration). The value of $\varphi=0.02$ is
nothing more than an educated guess, so its value can easily be wrong by a
factor of four or so. But the fact that the pressure scale height goes as
$T_{\mathrm{d}}^{1/2}\propto \varphi^{1/8}$ means that such an error only
affects the pressure scale height $h_p$ mildly.}
Assuming that the gas temperature is equal to the dust temperature, the pressure
scale height of the disk now follows from
\begin{equation}\label{eq-hp-afo-t}
h_p=\sqrt{\frac{k_BT_{\mathrm{d}}r^3}{\mu m_pGM_{*}}}
\end{equation}
with $k_B$ the Boltzmann constant, $m_p$ the proton mass, $G$ the
gravitational constant and $\mu=2.3$ the mean molecular weight in atomic
units. \removed{The stellar parameters are taken from Table \ref{tab-stellar-params}.}

We see from Table \ref{tab-gauss-params} that some rings are narrower than the
(estimated) pressure scale height $h_p$, while others are broader. This
comparison is important, because a long-lived pressure bump in the gas cannot be
radially narrower than about one pressure scale height. If it were, its
structure would be horizontally narrower than its vertical extent, which makes a
stable vertical hydrostatic equilibrium difficult to establish. \revised{Moreover, linear
stability analysis \citep[see][and Appendix
  \ref{sec-gas-ring-stability}]{2016ApJ...823...84O} shows that a Rossby wave
instability would be triggered, and the axial symmetry of the ring would be
lost.}

\revised{One can thus argue that,} if a thermal emission ring produced by the dust is
substantially narrower than $h_p$, then some kind of dust trapping must have
taken place. We can therefore conclude that we have strong evidence of dust trapping
operating in the rings in the disks around AS 209, Elias 24 and GW Lup. A
similar conclusion can be reached for the outer of the two high-contrast rings
in the disk around HD 163296, although the strong wing on the outer part makes
it harder to define the width unambiguously.  For the other rings dust trapping
is certainly not ruled out either, but would require further evidence.

As can be seen in Fig.~\ref{fig-obs-gaussfits}, for most rings the Gaussian
model fits the radial profile reasonably well, at least near the peak. The
largest relative deviation from a Gaussian shape can be seen in ring 1 of HD
163296. The peak of the profile is `pointier' than the best-fitting Gauss, and
the left flank steeper. On the other hand, the fitting window is much wider
than for the other ring profiles, and it remains close to the Gaussian fit
well into the wings.

In most rings the observed profiles rise above the Gaussian fit at some point in
the wings. This is particularly clear for the inner flanks of ring 1 of Elias 24
and ring 1 of HD 143006, as well as for the outer flanks of ring 2 of AS 209,
ring 2 of HD 163296, the ring of GW Lup and ring 2 of HD 143006. The excess
above the Gaussian gradually increases away from the peak of the Gaussian. The
profiles tend to Lorentzian shape in the flanks, but often asymmetrically.

For the double-ring objects HD 163296 and HD 143006,
Fig.~\ref{fig-obs-doublegauss} shows that the emission between the rings can
largely be explained by the overlapping Gaussians. In HD 143006 one could argue
that there is some excess (about twice as large as the the scatter along the
ring).

\section{Radial dust distribution}
The next step of our analysis is to investigate the spatial dust distribution
responsible for the ring emission.  \revised{As a first guess, we will assume that we can
ignore optical depth effects, and afterward we will consider optical depth
corrections.}

\subsection{Optically thin approximation}
\label{sec-opt-thin-analysis}
\revised{Let us first assume that the thermal emission of the dust is
optically thin.} The intensity profiles shown in Section
\ref{sec-data}, after deconvolution with the beam, are then linear maps of the
spatial distribution of dust, if we ignore any temperature gradients
or opacity gradients across these rings.\removed{In other words: the
deconvolved versions of the Gaussian fits of Section \ref{sec-gauss-fits} will
then directly translate in Gaussian models of the radial dust distribution.}
The conversion between the deconvolved observed intensity
profile $I_\nu^{\mathrm{dec}}(r)$ and the dust surface density profile
$\Sigma_d(r)$ is then
\begin{equation}\label{eq-optthin-conversion-linbright-sigmad}
  \Sigma_d^{\optthin{}}(r) =  \frac{I_\nu^{\mathrm{dec}}(r)}{\kappa_\nu^{\mathrm{abs}}\, B_\nu(T_d)} 
\end{equation}
where $T_d$ is the temperature of the dust, $\kappa_\nu^{\mathrm{abs}}$ is the
absorption opacity, and $B_\nu(T_d)$ is the Planck function.\removed{The superscript
``$\optthin$'' is to remind the reader that this is the surface density profile
under the assumption that the disk is optically thin.}

By replacing $I_\nu^{\mathrm{dec}}(r)$ with the Gaussian fit
$I_\nu^{\mathrm{gauss,dec}}(r)$ given by
Eq.~(\ref{eq-gauss-lin-br-temp-deconv}) we obtain the corresponding
$\Sigma_d^{\mathrm{gauss}}(r)$ from
Eq.~(\ref{eq-optthin-conversion-linbright-sigmad}). From this Gaussian
model we can derive the total dust mass trapped in the ring, ignoring
optical depth effects:
\begin{equation}\label{eq-dust-mass-estimate}
  M_d^{\optthin{}} = \int_0^\infty 2\pi r \Sigma_d^{\optthin{}}(r) dr \simeq
  \frac{(2\pi)^{3/2} r_0\, A\,\sigma}{\kappa_\nu^{\mathrm{abs}}\, B_\nu(T_d)}
\end{equation}
where we used the identity $A\,\sigma=A_{\mathrm{dec}}\,w_d$.

We use the DSHARP opacity model \citep{dsharp:birnstiel} which, for a grain
radius of $a=0.1\,\mathrm{cm}$ yields a dust opacity of
$\kappa_\nu^{\mathrm{abs}}(\lambda=0.125\,\mathrm{cm})=\kapabs{}\,\mathrm{cm}^2/\mathrm{g}$.
The resulting dust mass estimates are listed in Table \ref{tab-gauss-params}.

The main uncertainty lies in the opacity value $\kappa_\nu^{\mathrm{abs}}$. This
value depends on the grain size (or grain size distribution) as well as many
other factors including composition, grain shape and uncertainties in the method
of computation of the opacity. As shown in \citet{dsharp:birnstiel} the value of
$\kappa_\nu^{\mathrm{abs}}=\kapabs{}\,\mathrm{cm}^2/\mathrm{g}$ that we use here
can easily be wrong by \revised{a factor of $10$ upward or downward}, with
correspondingly large changes in the derived dust mass.


The other uncertainty is the dust temperature $T_d$, as we discussed before, but
this uncertainty is much less severe. For
Eq.~(\ref{eq-optthin-conversion-linbright-sigmad}) we need the corresponding
value of the Planck function $B_\nu(T_d)$, which is listed in \revised{column 12} in
Table \ref{tab-gauss-params}.

Given the amplitude of the deconvolved Gauss fit $A_{\mathrm{dec}}$ (see
Eq.~\ref{eq-gauss-lin-br-temp-deconv}), we can estimate the
optical depth $\tau_\nu^{\mathrm{peak}}$ of the ring at its peak at $r=r_0$:
\begin{equation}\label{eq-tau-estimate}
\tau_\nu^{\mathrm{peak}} = -\ln\left(1-\frac{A_{\mathrm{dec}}}{B_{\nu}(T_d)}\right)
\end{equation}
This estimate does not depend on the uncertain absorption opacity of the dust,
but it does depend on the dust temperature $T_d$, which depends on our
assumption of the flaring angle $\varphi$ through
Eq.~(\ref{eq-disk-temperature-model}). Fortunately, since
$T_d\propto\varphi^{0.25}$, we do not expect the temperature to be uncertain by
more than a factor of two, resulting in similar uncertainty in the optical depth
estimate.  The results are listed in \revised{column 15} of Table \ref{tab-gauss-params}.

\revised{We find optical depths of the order of
  $\tau_\nu^{\mathrm{peak}}\sim 0.2\cdots 0.5$, a surprisingly narrow range
  just below unity. For the case of HD 163296 there is independent evidence from
  the absorption of CO line emission from the back side of the disk that the
  optical depth in the two prominent rings is around \afterintrev{0.7}, as shown
  by \citet{dsharp:isella}. Evidently, the optically thin assumption is not
  entirely wrong, but not quite right either.}

\subsection{Optical depth corrections}
\label{sec-optical-depth-effects}
We have to verify how much the quantities we derive using the optically thin
assumption are affected by these optical depth effects. 
Let us assume that the dust has zero albedo.\removed{Then the inclusion of the
  effect of optical depth is simple.} We replace
Eq.~(\ref{eq-optthin-conversion-linbright-sigmad}) with the formal transfer
equation:
\begin{equation}\label{eq-simple-formal-rt}
  I_\nu^{\mathrm{dec}}(r) = \left(1-e^{-\tau_\nu(r)}\right)B_\nu(T_d)
\end{equation}
where $\tau_\nu(r)$ is the optical depth profile across the ring, and
\revised{we ignored any} background intensity, either from background clouds or
from the cosmic microwave background.

\removed{In interferometric observations a flat background is Fourier-filtered
  out, but in the radiative transfer it plays a role in the non-linear optical
  depth regime ($\tau_\nu\gtrsim 0.3$).  We assume, however, that the cloud
  background emission is negligible. The intensity of the cosmic background at
  $\lambda=0.125\,\mathrm{cm}$ is $I_\nu^{\mathrm{cmb}}=7\times
  10^{-3}\,\mathrm{Jy/arcsec}^2$, which is negligible compared to the mildly
  optically thick ring emission. We can therefore safely set
  $I_\nu^{\mathrm{bg}}=0$.}

To obtain the dust distribution we first compute $\tau_\nu(r)$
\begin{equation}\label{eq-tau-profile}
\tau_\nu(r) = -\ln\left(1-\frac{I_\nu^{\mathrm{dec}}(r)}{B_\nu(T_d)}\right)
\end{equation}
\removed{which is identical to Eq.~(\ref{eq-tau-estimate}), but now for the entire
profile instead of just for the peak.} The profile for $\Sigma_d(r)$ now follows
from
\begin{equation}\label{eq-sigmadust-from-tau-and-kappa}
\Sigma_d(r) = \frac{\tau_\nu(r)}{\kappa_\nu^{\mathrm{abs}}}
\end{equation}
The problem is, of course, that it is not straightforward to deconvolve the
observed $I_\nu(r)$ profile if the underlying $I_\nu^{\mathrm{dec}}(r)$ is not a
Gaussian.

\revised{Strong optical depth effects should lead to flat-topped radial ring
  profiles. The radial ring profiles of this paper do not appear to show such
  flat-topped shapes, which means that the rings in our sample cannot be highly
  optically thick. This is in agreement with our estimates of
  $\tau_\nu^{\mathrm{peak}}$ being of the order 0.2$\cdots$0.5.}

\revised{At the moderate optical depths of our rings, the optical depth
  correction mainly leads to an upward correction of the derived dust surface
  density $\Sigma_d(r)$ and the corresponding dust masses $M_d$. As one can see in
  Table \ref{tab-gauss-params}, this effect is relatively minor, in particular
  compared to the uncertainties of the opacity model.}

\begin{figure}
\centerline{\includegraphics[width=0.47\textwidth]{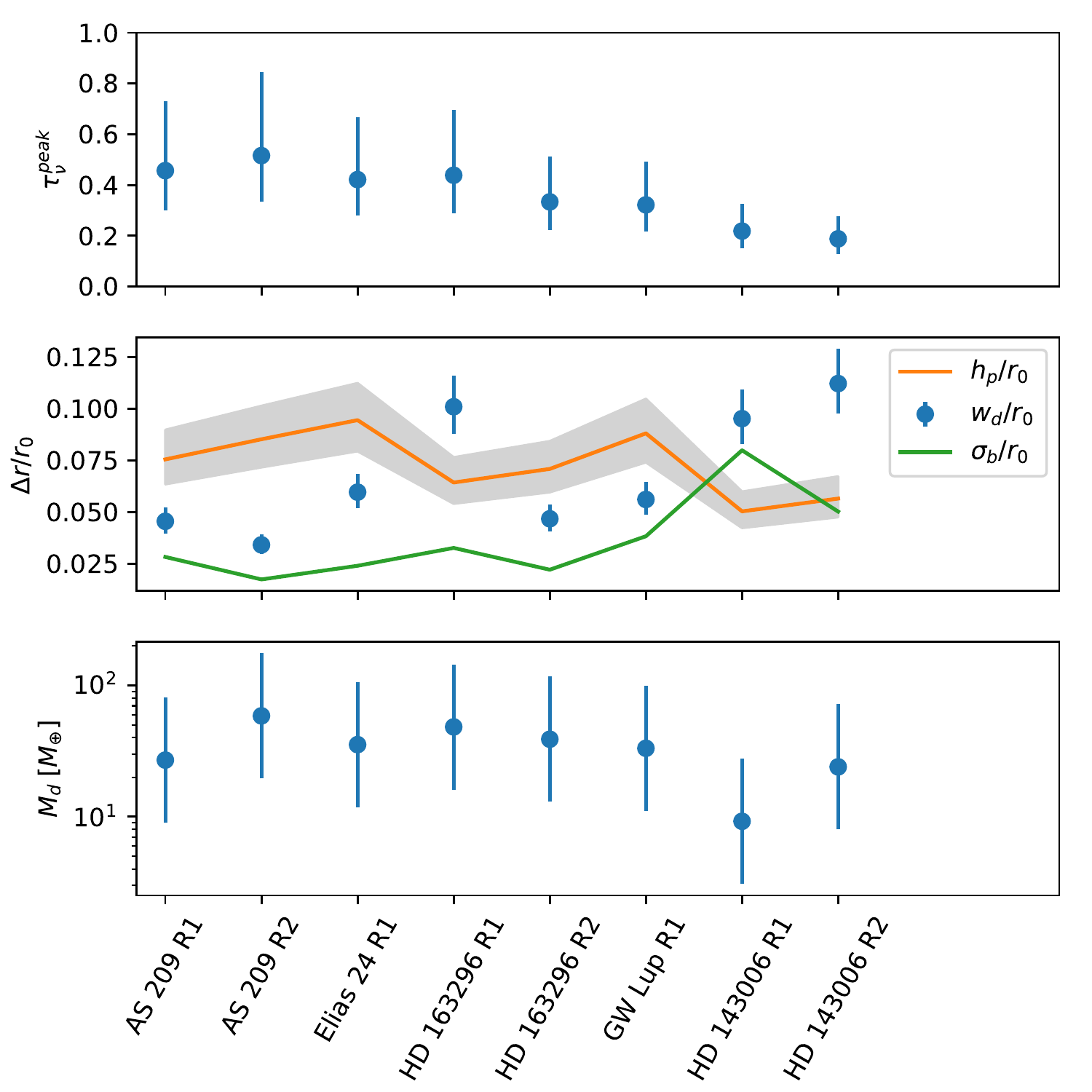}}
\caption{\label{fig-summary-1}Summary of the numbers resulting from the Gaussian
  fitting of the radial profiles of the rings, as listed in Table
  \ref{tab-gauss-params}. Top: The optical depth of the ring at the peak of the
  intensity ($\tau_\nu^{\mathrm{peak}}$). Middle: The relative size, in units of
  the ring radius $r_0$, of the dust ring width $w_d$, the pressure scale height
  $h_p$ and the standard deviation beam size $\sigma_b$.}
\end{figure}

The most important results we obtained so far are summarized in
Fig.~\ref{fig-summary-1}. The uncertainties of $\tau_\nu^{\mathrm{peak}}$ and
$h_p$ are both estimated from an estimated uncertainty of the dust temperature
$T_{\mathrm{d}}$ through Eqs.~(\ref{eq-tau-profile}, \ref{eq-hp-afo-t}), because
this is by far the largest source of uncertainty. We assume a factor of (0.25,4)
uncertainty of the irradiating flux, yielding roughly an uncertainty of
$(\sqrt{0.5},\sqrt{2})$ in $T_{\mathrm{d}}$. The uncertainty in the dust mass is
estimated from the uncertainty in the opacity through
Eq.~(\ref{eq-sigmadust-from-tau-and-kappa}).

\section{The rings as dust traps}
\label{sec-rings-as-dust-traps}
The hypothesis we are now going to test is that the
rings are caused by dust trapping in axisymmetric pressure bumps. For simplicity
we will assume that the radial gas pressure profile is fixed in time, and there
is no back-reaction of the dust onto the gas. The pressure bump is assumed to be
so strong that the dust trapping in these rings is perfect: no dust
escapes. We then expect that the dust distribution finds an equilibrium between
dust drift and turbulent spreading. \removed{Low turbulence will lead to narrower dust
rings than high turbulence. For simplicity we shall assume the gas pressure bump
to have a Gaussian radial profile with its peak at radius $r_0$ and width $w\ll
r_0$. This problem can be solved analytically.}

\subsection{Model}
\label{sec-model-dusttrap-rings}
Consider the following radial Gaussian profile for the
pressure at the disk midplane:
\begin{equation}\label{eq-gaussian-pressure-bump}
p(r) = p_0 \exp\left(-\frac{(r-r_0)^2}{2w^2}\right)
\end{equation}
\revised{where $w$ is the width, and $p_0$ is the pressure at the peak of the
  pressure bump, located at $r=r_0$. The width has to obey $w\ge h_p$ to ensure
  stability (see Appendix \ref{sec-gas-ring-stability}).}

\revised{The equilibrium between radial drift and radial mixing leads to the
  following radial distribution of the dust (see Appendix
  \ref{sec-steady-state-analytic-trap-model} for the derivation):}
\begin{equation}\label{eq-analytic-sol-radial-trapping-summary}
\Sigma_{\mathrm{d}}(r) = \Sigma_{\mathrm{d0}} \exp\left(-\frac{(r-r_0)^2}{2w_{\mathrm{d}}^2}\right)
\end{equation}
where
\begin{equation}\label{eq-wd-afo-w-psi}
  w_{\mathrm{d}} = w\,\left(1+\psi^{-2}\right)^{-1/2}
\end{equation}
with $\psi$ given by
\begin{equation}\label{eq-psi-afo-alpha-sc-st}
\psi = \sqrt{\frac{\alpha_{\mathrm{turb}}}{\mathrm{Sc}\,\mathrm{St}}}
\end{equation}
Here $\mathrm{St}$ is the Stokes number of the dust particles
(Eq.~\ref{eq-definition-stokes-number}), $\mathrm{Sc}$ is the Schmidt number of
the turbulence in the gas \afterintrev{(the ratio between turbulent viscosity
  and turbulent diffussivity)}, and $\alpha_{\mathrm{turb}}$ is the usual
turbulence parameter. Note that this solution is for a single grain size.

For large grains and/or weak turbulence one finds $\psi\ll 1$, which leads to
$w_{\mathrm{d}}\ll w$. In this case the dust is strongly trapped near the peak
of the pressure bump. The opposite is the case for small grains and/or strong
turbulence, for which one gets $\psi\gg 1$, which leads to
$w_{\mathrm{d}}\rightarrow w$. In this case the trapping is very weak and the
dust-to-gas ratio within the pressure bump stays nearly constant.

\afterintrev{It is interesting to note that this parameter $\psi$ also
  determines the degree of vertical settling of the same dust:}
\begin{equation}\label{eq-hd-afo-hp-psi}
h_{\mathrm{d}} = h_p \,\left(1+\psi^{-2}\right)^{-1/2}
\end{equation}
\afterintrev{In other words: dust particles that are radially trapped in a
  narrow ring are also vertically settled. This does not mean, however, that
  dust that is not settled can always radially drift through any dust trap.  In
  fact: even for $\psi\gg 1$ our model still assumes that all the dust remains
  trapped, albeit in the far wings of the Gaussian pressure trap. This has
  relevance for dust trapping in the edges of planetary gaps, which we will
  discuss in Section \ref{sec-planet-gap}.}

Eq.~(\ref{eq-analytic-sol-radial-trapping-summary}) has only \afterintrev{three} parameters:
$\Sigma_{\mathrm{d0}}$, $w_{\mathrm{d}}$ and $r_0$. As we have shown in Sections
\ref{sec-gauss-fits}, \ref{sec-opt-thin-analysis} and
\ref{sec-optical-depth-effects}, \afterintrev{all three} parameters can be extracted from the
observations. The main uncertainty lies in $\Sigma_{\mathrm{d0}}$, due to the
uncertainty in the dust opacity. The values of $w_{\mathrm{d}}$ for the rings in
our sample can be directly taken from Table \ref{tab-gauss-params}.

The width of the dust ring $w_{\mathrm{d}}$ is physically set by
$\alpha_{\mathrm{turb}}$, $\mathrm{Sc}$, $\mathrm{St}$ and $w$ through the above
equations. We therefore
have one observational value for four unknown parameters. This is heavily
degenerate. All we can do is to test if the measured value of
$w_{\mathrm{d}}$ is consistent with expected values of $\alpha_{\mathrm{turb}}$,
$\mathrm{Sc}$, $\mathrm{St}$ and $w$.

\subsection{Limits to $\alpha_{\mathrm{turb}}$, $\mathrm{Sc}$, $\mathrm{St}$ and $w$}
\label{sec-ranges-of-params}
Reasonable values of $\alpha_{\mathrm{turb}}$, $\mathrm{Sc}$, $\mathrm{St}$ and
$w$ obey certain restrictions. First of all, the Schmidt number $\mathrm{Sc}$
is merely a way to relate the turbulent viscosity with the turbulent mixing. If
we do not strive to learn about the turbulent viscosity, and instead are
satisfied with learning only about the turbulent mixing, then we are only
interested in the combination $\alpha_{\mathrm{turb}}/\mathrm{Sc}$. For
simplicity we set $\mathrm{Sc}=1$, \afterintrev{which is a reasonable
value \citep{2005ApJ...634.1353J}.}

The value of the turbulence parameter $\alpha_{\mathrm{turb}}$ is usually
considered to be between $10^{-6}\lesssim \alpha_{\mathrm{turb}}\lesssim
10^{-2}$.

The width of the pressure bump cannot be smaller than about a pressure
scale height, but also not smaller than the width of the dust ring.
Therefore $w_{\mathrm{min}}=\mathrm{max}(h_p,w_{\mathrm{d}})$.
In the case of the double rings (AS 209, HD 163296 and HD 143006),
the full-width-at-half-maximum $2.355\,w$ should not exceed the radial
separation of the rings. For the two single ring sources we take the
\afterintrev{distance from the peak to the} deepest
point of the gap to the inside of the ring as the upper limit on the
half-width-at-half-maximum $1.178\,w$. These lower and upper limits on $w$
are listed in Table \ref{tab-ring-model-limits}.

The Stokes number $\mathrm{St}$ can be any value. But it is directly related to
the grain size $a$ and the gas density $\rho_{\mathrm{g}}$, where the gas
density is directly related to the gas surface density $\Sigma_{\mathrm{g}}$ via
$\Sigma_{\mathrm{g}}=\sqrt{2\pi}h_p\rho_{\mathrm{g}}$. If we have observational
constraints on \afterintrev{the grain size} $a_{\mathrm{grain}}$ and a good estimate of the gas surface density
$\Sigma_{\mathrm{g}}$, then we can eliminate this uncertainty, and we are left
with two unknown parameters ($\alpha_{\mathrm{turb}}$ and $w$) for one
measurement ($w_{\mathrm{d}}$).  Unfortunately, while estimating $a_{\mathrm{grain}}$ from
observations may be doable, it is far more difficult to estimate
$\Sigma_{\mathrm{g}}$. Standard disk gas mass estimates are of limited use, as
they are based on measuring the dust mass and multiplying it by the estimated
gas-to-dust ratio. Since we are testing the hypothesis of dust trapping, we
cannot assume a standard gas-to-dust ratio.

One can, however, set an upper bound on $\Sigma_{\mathrm{g}}$ by demanding that
the disk is gravitationally stable, i.e.\ that the Toomre parameter obeys
\begin{equation}\label{eq-sigmag-upper-lim}
Q_{\mathrm{Toomre}}\equiv \frac{c_s\Omega_K}{\pi G \Sigma_{\mathrm{g}}} >2
\end{equation}
Otherwise non-axisymmetric features, such as spiral arms, would develop
\citep[see e.g.][]{2016ARA&A..54..271K}, which would also be seen in the
continuum emission. Here $c_s=\sqrt{k_BT_{\mathrm{g}}/\mu m_p}$ is the
isothermal sound speed (with $T_{\mathrm{g}}$ being the gas temperature),
$\Omega_K=\sqrt{GM_{*}/r^3}$ is the Kepler frequency, $G$ is the gravitational
constant, and $\Sigma_{\mathrm{g}}$ the gas surface density. Taking the disk
midplane temperature from Table \ref{tab-gauss-params}, which was calculated
using Eq.~(\ref{eq-disk-temperature-model}), we can compute the upper limits on
$\Sigma_{\mathrm{g}}$ for all of the rings. The results are listed in Table
\ref{tab-ring-model-limits} as $\Sigma_{\mathrm{g,max}}$.

\begin{deluxetable*}{ccccccccccccc}[b!]
\tablecaption{Limits on the free parameters of the dust
  trapping model.\label{tab-ring-model-limits}}
\tablecolumns{13}
\tablewidth{0pt}
\tablehead{
\colhead{Source} &
\colhead{Ring} &
\colhead{Name} &
\colhead{$w_{\mathrm{min}}$} &
\colhead{$w_{\mathrm{max}}$} &
\colhead{$\Sigma_{\mathrm{g,min}}$} &
\colhead{$\Sigma_{\mathrm{g,max}}$} &
\colhead{$a_{\mathrm{max}}$} &
\colhead{$\mathrm{St}_{(a=0.02\,\mathrm{cm})}$} &
\colhead{$w_d/w$} &
\colhead{$\alpha/\mathrm{St}$} &
\colhead{$\alpha/\mathrm{St}$} &
\colhead{$\alpha_{\mathrm{exmp}}$} \\
\colhead{} &
\colhead{} &
\colhead{} &
\colhead{$[\mathrm{au}]$} &
\colhead{$[\mathrm{au}]$} &
\colhead{$[\mathrm{g}/\mathrm{cm}^2]$} &
\colhead{$[\mathrm{g}/\mathrm{cm}^2]$} &
\colhead{$[\mathrm{cm}]$} &
\colhead{(for $\Sigma_{\mathrm{g,max}}$)} &
\colhead{(for $w_\mathrm{max}$)} &
\colhead{(for $w_\mathrm{max}$)} &
\colhead{(for $w_\mathrm{min}$)} &
\colhead{} \\ 
\colhead{\colsrcname{}} &
\colhead{\colring{}} &
\colhead{\colringname{}} &
\colhead{\colwmin} &
\colhead{\colwmax} &
\colhead{\colsiggmin} &
\colhead{\colsiggmax} &
\colhead{\colamax} &
\colhead{\colstexmp{}} &
\colhead{\colwdw} &
\colhead{\colastwmax} &
\colhead{\colastwmin} &
\colhead{\colaexmp} 
}
\startdata
AS 209     & 1 & B74  &   5.6 & 19.6 & 2.3e-01 & 1.6e+01 &  4.7 & 3.2e-03 &  0.17 & 3.1e-02 & 5.7e-01 & 9.9e-05\\
AS 209     & 2 & B120 &  10.3 & 19.6 & 2.6e-01 & 6.9e+00 &  1.2 & 7.6e-03 &  0.21 & 4.6e-02 & 1.9e-01 & 3.5e-04\\
Elias 24   & 1 & B77  &   7.2 & 17.1 & 2.1e-01 & 1.8e+01 &  5.8 & 3.0e-03 &  0.27 & 7.7e-02 & 6.6e-01 & 2.3e-04\\
HD 163296  & 1 & B67  &   6.8$^{*}$ & 13.8 & 2.2e-01 & 4.0e+01 & 15.4 & 1.3e-03 &  0.50 & 3.3e-01 & -- & 4.2e-04\\
HD 163296  & 2 & B100 &   7.1 & 13.8 & 1.7e-01 & 2.0e+01 &  9.5 & 2.6e-03 &  0.34 & 1.3e-01 & 7.7e-01 & 3.3e-04\\
GW Lup     & 1 & B85  &   7.5 &  9.9 & 1.6e-01 & 7.8e+00 &  2.7 & 6.7e-03 &  0.48 & 3.1e-01 & 6.8e-01 & 2.1e-03\\
HD 143006  & 1 & B41  &   3.9$^{*}$ & 10.1 & 1.1e-01 & 7.5e+01 & 60.6 & 7.0e-04 &  0.39 & 1.8e-01 & -- & 1.2e-04\\
HD 143006  & 2 & B65  &   7.3$^{*}$ & 10.1 & 9.4e-02 & 3.4e+01 & 30.8 & 1.6e-03 &  0.72 & 1.1e+00 & -- & 1.7e-03\\
\enddata
\tablecomments{Columns 1 to 3 are the same as in Table \ref{tab-gauss-params}.
  \colwmin{} Lower limit to the pressure bump width $w_{\mathrm{min}}$ (for
  $w_d\le h_p$ this is $h_p$; for $w_d>h_p$, marked with the symbol $^{*}$, this is $w_d$).
  \colwmax{} Upper limit to the pressure bump width $w_{\mathrm{max}}$,
  derived from the separation between the
  rings (for AS 209, HD 163296 and HD 143006) or from the separation of the ring
  to the nearest minimum (for Elias 24 and GW Lup). \colsiggmin{} Lower limit on the gas
  surface density $\Sigma_{\mathrm{g}}$ derived by demanding
  $\Sigma_{\mathrm{g}}\gtrsim \Sigma_{\mathrm{d}}$. Note that this involves the
  uncertainty in $\Sigma_{\mathrm{d}}$ due to the uncertainty of the dust
  opacity model. \colsiggmax{} Upper
  limit on the gas surface density derived from demanding that the gas disk is
  gravitationally stable. \colamax{} Maximum grain
  size $a_{\mathrm{max}}$ for which the derived dust surface density (based on the DSHARP opacity
  model) together with the gas surface density remain gravitationally
  stable. \colstexmp{} Example value of the Stokes number $\mathrm{St}$
  for grains with a radius of $0.02\,\mathrm{cm}$. \colwdw{} Estimate of the degree of
  dust trapping given by the ratio  $w_d/w$ (assuming that
  $w=w_{\mathrm{max}}$). The smaller this number is, the stronger the dust trapping.
  \colastwmax{} Value of $\alpha/\mathrm{St}$ derived for the widest gas bump.
  \colastwmin{} Value of $\alpha/\mathrm{St}$ derived for the narrowest gas bump.
  \colaexmp{} Example value of $\alpha_{\mathrm{turb}}$, computed for 
  $w=w_{\mathrm{max}}$, $\Sigma_{\mathrm{g}}=\Sigma_{\mathrm{g,max}}$ and
  $a=0.02\,\mathrm{cm}$.}
\end{deluxetable*}

One can estimate a lower limit to the gas density by demanding that the gas
surface density must be at least as large as the dust surface density, since
dust trapping is unlikely to achieve a larger concentration of dust than
that. For the dust surface density we use
Eq.~(\ref{eq-sigmadust-from-tau-and-kappa}) at $r=r_0$, with
$\tau_\nu(r_0)=\tau_\nu^{\mathrm{peak}}$ from Table \ref{tab-gauss-params}. By
demanding that
\begin{equation}\label{eq-sigmag-lower-lim}
\Sigma_{\mathrm{g}}\gtrsim \Sigma_{\mathrm{d}}(r_0) = \frac{\tau_\nu^{\mathrm{peak}}}{\kappa_\nu^{\mathrm{abs}}}
\end{equation}
and using our standard opacity of
$\kappa_\nu^{\mathrm{abs}}=\kapabs{}\,\mathrm{cm}^2/\mathrm{g}$ we arrive at
values for $\Sigma_{\mathrm{g,min}}(r=r_0)$ listed in Table
\ref{tab-ring-model-limits} as $\Sigma_{\mathrm{g,min}}$.

It is likely that even for larger values of the gas surface density the dust-gas
mixture becomes unstable to the streaming instability and other types of
instabilities, because the dust will likely settle to the midplane, increasing
the ratio $\rho_{\mathrm{d}}/\rho_{\mathrm{g}}$. We can quantify this. For a
given ratio $\alpha_{\mathrm{turb}}/\mathrm{St}$, we can compute the ratio
$h_d/h_p$ from Eqs.~(\ref{eq-psi-afo-alpha-sc-st}, \ref{eq-hd-afo-hp-psi}),
which tells us how strongly the dust is settled. The new (and more stringent)
lower limit to the gas density is then
$\Sigma_{\mathrm{g,min}}^{\mathrm{sett}}=(h_p/h_d)\,\Sigma_{\mathrm{g,min}}$.
\removed{ The
values of $\Sigma_{\mathrm{g,min}}^{\mathrm{sett}}$ are not listed in the table,
because they are parameter-dependent. But we can obtain some fiducial values. If
we choose the highest possible $\alpha_{\mathrm{turb}}/\mathrm{St}$ (the one for
$w_{\mathrm{min}}$ in Table \ref{tab-ring-model-limits}) we obtain
$\Sigma_{\mathrm{g,min}}^{\mathrm{sett}}$ only moderately increased with respect
to $\Sigma_{\mathrm{g,min}}$. \revised{For example, ring 2 of AS 209 with
$h_d/h_p=0.4$ for $\alpha_{\mathrm{turb}}/\mathrm{St}=0.19$ (see column
  12 of Table \ref{tab-ring-model-limits}), yields
$\Sigma_{\mathrm{g,min}}^{\mathrm{sett}} \simeq
2.5\,\Sigma_{\mathrm{g,min}}=0.7\,\mathrm{g/cm}^2$}. If we choose the lowest
possible $\alpha_{\mathrm{turb}}/\mathrm{St}$ (\revised{column 11 of}
Table \ref{tab-ring-model-limits}), then for \revised{the same ring
  we obtain $h_d/h_p=0.21$, yielding
$\Sigma_{\mathrm{g,min}}^{\mathrm{sett}} \simeq 4.8\,\Sigma_{\mathrm{g,min}}
=1.24\,\mathrm{g/cm}^2$.}}

If the grains are much larger than $\lambda/(2\pi)\simeq 0.02\,\mathrm{cm}$,
the opacity drops and the resulting dust surface density estimate increases,
also yielding larger values of $\Sigma_{\mathrm{g,min}}$.
Along this line of thinking one can compute the largest grain radius for which
$\Sigma_{\mathrm{g,min}}<\Sigma_{\mathrm{g,max}}$, i.e.\ for which the
$\Sigma_{\mathrm{g,min}}$ is consistent with $Q_{\mathrm{Toomre}}>2$. This gives a lower limit to
the dust opacity $\kappa_\nu^{\mathrm{abs}}$ and, as a result, an upper limit to
the grain size. Given that the total surface density is then twice the gas
surface density (the dust contributing the other half), we have to introduce a
factor of 2. The condition on the opacity is then:
\begin{equation}
\kappa_\nu^{\mathrm{abs}} \gtrsim \frac{2\tau_\nu^{\mathrm{peak}}}{\Sigma_{\mathrm{g,max}}}
\end{equation}
We now use the DSHARP opacity model \citep{dsharp:birnstiel} to translate this
$\kappa_\nu^{\mathrm{abs}}$ into a grain radius. We arrive at values of
centimeters to half a meter (Table \ref{tab-ring-model-limits}). These are
conservative limits, with real values likely to be substantially smaller.
\revised{Indeed, in the next Subsection we will derive, from the values of
  $\alpha_{\mathrm{turb}}/\mathrm{St}$ in Table \ref{tab-ring-model-limits},
  much more stringent upper limits on the grain size.}

\subsection{Application to the observed rings}
\label{sec-application-to-rings}
We now apply the model of Subsection \ref{sec-model-dusttrap-rings} with the
limits on the parameter ranges derived in Subsection \ref{sec-ranges-of-params}
to the observed ring widths $w_d$ listed in Table
\ref{tab-gauss-params}. \afterintrev{The goal is to see which constraints the
  observations can put on the physics of the observed rings of this paper.}

From an assumed value of $w$ and the measured value $w_{\mathrm{d}}$ we can
directly compute the ratio $\alpha_{\mathrm{turb}}/\mathrm{St}$
\begin{equation}\label{eq-alphaSt-equation}
\frac{\alpha_{\mathrm{turb}}}{\mathrm{St}} \equiv \psi^{2} = \left[\left(\frac{w}{w_{\mathrm{d}}}\right)^2-1\right]^{-1}
\end{equation}
where we used Eqs.~(\ref{eq-wd-afo-w-psi}, \ref{eq-psi-afo-alpha-sc-st}), and
set $\mathrm{Sc}=1$. We will consider two choices of $w$: the $w_{\mathrm{min}}$ and
$w_{\mathrm{max}}$ from Table \ref{tab-ring-model-limits}.

For the choice $w=w_{\mathrm{max}}$ (the widest possible pressure bump) the dust
rings are all narrower than the gas rings: $w_{\mathrm{d}}<w$, as can be seen
from the $w_d/w$ column in Table \ref{tab-ring-model-limits}. This implies
that, under the assumption that $w=w_{\mathrm{max}}$, the dust trapping is
operational. The ratio $w_d/w$ gives an indication of the degree of dust
trapping: the smaller this value is, the closer the dust has drifted to the peak
of the pressure bump before turbulent mixing halts further narrowing of the dust
ring. The strength of the turbulence for this case is given by the
$\alpha_{\mathrm{turb}}/\mathrm{St}$ column for $w=w_{\mathrm{max}}$ in Table
\ref{tab-ring-model-limits}.

One important result from this analysis is that, although these rings are the
narrowest that have been observed so far, the ratio $w_d/w$ is never smaller
than 17\%, usually subtantially larger. This means that in all these rings
turbulence prevents the dust from forming even narrower dust rings.  Perhaps
this is self-induced turbulence due to the large $\Sigma_{d}/\Sigma_g$ ratio in
this dust trap. Or it could mean that the dust is still not yet in drift-mixing
equilibrium, which would require the grains to be very small (i.e.\ have a very
low value of $\mathrm{St}$). In Section \ref{sec-planet-gap} we will discuss an
example of the latter scenario.

For the choice $w=w_{\mathrm{min}}$
(the narrowest possible pressure bump) we can only use
Eq.~(\ref{eq-alphaSt-equation}) for the rings for which
$w_{\mathrm{d}}<h_p$. The reason is that for those rings with
$w_{\mathrm{d}}>h_p$ (marked with a $^{*}$ in Table \ref{tab-ring-model-limits})
the minimal pressure bump width is $w_{\mathrm{min}}=w_{\mathrm{d}}$, and the
dust ring is as wide as the pressure bump, implying that dust trapping is weak
or non-operational. Any increase of $\alpha_{\mathrm{turb}}/\mathrm{St}$ will
keep $w_{\mathrm{d}}=w_{\mathrm{min}}$, so one cannot derive any value for
$\alpha_{\mathrm{turb}}/\mathrm{St}$. But for other rings (those not marked
with *) we can compute
$\alpha_{\mathrm{turb}}/\mathrm{St}$. The resulting values for both choices of
pressure bump width are given in Table \ref{tab-ring-model-limits}, \revised{columns 11
  and 12}. They can be
understood as the lower and upper limit on $\alpha_{\mathrm{turb}}/\mathrm{St}$.

\afterintrev{We conclude that for those rings not marked with the ${*}$-symbol
  in Table \ref{tab-ring-model-limits}, our data is clear proof of dust trapping
  occurring. For the rings marked with $*$ the narrowness of the dust ring can
  also be explained simply by the narrowness of the underlying gas ring without
  the need for dust trapping, although it does not exclude dust trapping either.}

The next task is to convert from Stokes number $\mathrm{St}$ to grain radius
$a_{\mathrm{grain}}$. The Epstein regime is valid for grain sizes of the order of millimeters
or centimeters, in which case $a_{\mathrm{grain}}$ and $\mathrm{St}$ are related by 
\begin{equation}\label{eq-agrain-to-stokes}
\mathrm{St} = \frac{\pi}{2}\frac{\xi_{\mathrm{dust}}a_{\mathrm{grain}}}{\Sigma_{\mathrm{g}}}
\end{equation}
where $\Sigma_{\mathrm{g}}$ is the gas surface density and $\xi_{\mathrm{dust}}$
is the material density of the dust grains. For the DSHARP opacity
model \citep{dsharp:birnstiel} the average material density of the dust
aggregates is $\xi_{\mathrm{dust}}\simeq \xidust{}\,\mathrm{g}/\mathrm{cm}^3$.

To get a feeling for the results, let us choose the grain size to be
$a_{\mathrm{grain}}=0.02\,\mathrm{cm}$, which corresponds to $\lambda/2\pi$ for
$\lambda=0.125\,\mathrm{cm}$ (the wavelength of ALMA band 6).  The corresponding
Stokes numbers, for the most massive possible gas disk
($\Sigma_{\mathrm{g}}=\Sigma_{\mathrm{g,max}}$), are listed in Table
\ref{tab-ring-model-limits}, \revised{column 9}. This then allows us to convert the value of
$\alpha_{\mathrm{turb}}/\mathrm{St}$ into a value of
$\alpha_{\mathrm{turb}}$, which we shall call $\alpha_{\mathrm{exmp}}$,
  indicating that it is an example value for a particular choice of
$a_{\mathrm{grain}}$. For the case $w=w_{\mathrm{max}}$ this leads to values
$\alpha_{\mathrm{exmp}}= 10^{-4}\ldots\mathrm{few}\,\times 10^{-3}$, listed in
Table \ref{tab-ring-model-limits}, \revised{column 13}.

These low values of $\alpha_{\mathrm{turb}}$ are consistent with the low values
or upper limits reported recently \citep{2016ApJ...816...25P,
  2018ApJ...856..117F}. However, it has to be kept in mind that the values of
$\alpha_{\mathrm{turb}}=\alpha_{\mathrm{exmp}}$ were derived for an extremal
choice of parameters: $w=w_{\mathrm{max}}$,
$\Sigma_{\mathrm{g}}=\Sigma_{\mathrm{g,max}}$, and only for grain radius
$a=0.02\,\mathrm{cm}$. For a smaller value of $w$, a lower value of
$\Sigma_{\mathrm{g}}$, or larger grains, the computed value of
$\alpha_{\mathrm{turb}}$ will increase. If we take ring 1 of AS 209 as an
example, and take $w=w_{\mathrm{min}}$, we see from Table
\ref{tab-ring-model-limits} that $\alpha/\mathrm{St}=0.57$. Using
\revised{$\Sigma_{\mathrm{g}}=\Sigma_{\mathrm{g,min}}$} (but keep $a_{\mathrm{grain}}=0.02\,\mathrm{cm}$)
we get $\mathrm{St}=0.23$ \revised{from Eq.~(\ref{eq-agrain-to-stokes})}, yielding $\alpha_{\mathrm{turb}}=0.13$.
\revised{This is much higher than the value of $\alpha_{\mathrm{exmp}}$, and it}
demonstrates that it is hard to set a true upper limit on $\alpha_{\mathrm{turb}}$
from these observations. 

Can we derive a {\em lower} limit to $\alpha_{\mathrm{turb}}$? This depends on
whether we have information about the grain size. The value of
$\alpha_{\mathrm{exmp}}$ \revised{is the smallest possible value of
  $\alpha_{\mathrm{turb}}$ consistent with the data, for an assumed grain size
  of $a=0.02\,\mathrm{cm}$.} Since Eq.~(\ref{eq-agrain-to-stokes}) shows that
$a_{\mathrm{grain}}$ and $\mathrm{St}$ are linearly related, \revised{we
can generalize this to the smallest possible value of
$\alpha_{\mathrm{turb}}$ consistent with the data, for any given
grain size $a_{\mathrm{grain}}$:
\begin{equation}\label{eq-limit-on-turbulence}
\alpha_{\mathrm{turb}}\ge \left(\frac{a_{\mathrm{grain}}}{0.02\,\mathrm{cm}}\right)\;\alpha_{\mathrm{exmp}}
\end{equation}
}
With the values of
$\alpha_{\mathrm{exmp}}$ listed in Table \ref{tab-ring-model-limits} this shows
that, even for disks so massive that they are nearly gravitationally unstable,
we can exclude the combination of very low $\alpha_{\mathrm{turb}}\ll 5\times
10^{-4}$ and very large grains $a_{\mathrm{grain}}\gg 0.1\,\mathrm{cm}$ for all
the rings of our sample. In many of the rings this constraint is much more
strict (i.e.~toward smaller grains and/or stronger turbulence).

To obtain estimates of the grain size we need spectral information. At present
we have only the high resolution data for band 6, so we do not yet have
information about the radial profile of the spectral slope. But in several
recent observations of the spectral index across ringed disks
\citep{2015ApJ...808L...3A, 2016ApJ...829L..35T, 2018ApJ...852..122H} one
clearly sees that $\alpha_{\mathrm{spec}}$ varies across these rings, being
closer to $2$ at the ring center and substantially larger between the
rings. This makes sense in terms of the dust trapping scenario in which we
expect larger grains to be trapped more efficiently (and thus dominate the peak
of the ring) than smaller grains, because the smaller grains will be more
subject to turbulent mixing. \afterintrev{It is clear that we need such data to
  be able to constrain $a_{\mathrm{grain}}$, and then, via
  Eq.~(\ref{eq-limit-on-turbulence}), set limits on the turbulence.}

\subsection{Including a grain size distribution}
\label{sec-model-with-grain-size-distribution}
So far we have only looked at a single grain size, for which the solution is a
Gaussian radial grain distribution centered around the point of zero gas
pressure gradient. The model fits fairly well the near-Gaussian profiles that we
observe. However, in several rings we find a deviation from Gauss in the form of
an excess emission in the wings of the profile. \revised{Could this be a result
  of a grain size distribution?}
\removed{One explanation is, of course, that the gas pressure profile is not Gaussian. We
  will investigate this case in Section \ref{sec-planet-gap}.
  Here we investigate, however, the scenario in which the emission in the wings is
caused by smaller dust grains that are less well trapped, i.e.\ have a larger
$w_d$.}
\revised{To find out, let us apply our model to the following powerlaw
  size distribution:}
\begin{equation}
m(a)\frac{dN}{d\ln a} = \frac{dM}{d\ln a} \propto a^{p}
\end{equation}
where $a$ is the grain size, $m(a)$ the corresponding grain mass, $N$ the
cumulative particle number and $M$ the cumulative dust mass. The parameter $p$
is the size distribution powerlaw coefficient, and it is $p=1/2$ for the usual
MRN distribution (this corresponds to $dN/da\propto a^{p-4} = a^{-3.5}$). We also need to
define limits $a_{\mathrm{min}}$ and $a_{\mathrm{max}}$. \removed{The size distribution
is then normalized such that its integral over $d\ln(a)$ is the total dust mass
$M$.} The radial surface density solution,
Eq.~(\ref{eq-analytic-sol-radial-trapping}), then becomes:
\begin{equation}\label{eq-analytic-sol-radial-trapping-sizedistr}
\frac{d\Sigma_{\mathrm{d}}(r)}{d\ln a} = \frac{1}{(2\pi)^{3/2}r_0
    w_{\mathrm{d}}(a)}\frac{dM}{d\ln a}
\exp\left(-\frac{(r-r_0)^2}{2w_{\mathrm{d}}(a)^2}\right)
\end{equation}
\revised{At each radius $r$ the local size distribution is different from other
  radii, with larger grains dominating near $r=r_0$ and smaller grains
  dominating in the wings.}

To demonstrate the effect we will try to apply this multi-size dust trapping
model to ring 1 of AS 209. We set the gas ring width to $w=19.6\,\mathrm{au}$,
and gas surface density to $\Sigma_g=16\,\mathrm{g}\,\mathrm{cm}^{-2}$,
i.e.\ the maximum $w$ and $\Sigma_g$ as listed in Table
\ref{tab-ring-model-limits}.  We set $\alpha_{\mathrm{turb}}=1.1\times 10^{-3}$,
$p=1/2$ (MRN slope), $a_{\mathrm{min}}=10^{-2}\,\mathrm{cm}$ and
$a_{\mathrm{max}}=1\,\mathrm{cm}$. We take 10 grain size bins logarithmically
spaced in $a$. For the rest we take the same parameters
as listed in Table \ref{tab-gauss-params}. Since the relation between the
observed emission and the underlying dust mass is different if we take a size
distribution, we adjust the dust mass such that the model yields a peak optical
depth equal to the $\tau_\nu^{\mathrm{peak}}$ value from Table
\ref{tab-gauss-params}. \removed{To keep things simple, we will not worry about the
$1-e^{-\tau}$ optical depth effects, assuming the relation between
$I_\nu$ and $\tau_\nu$ to be linear. But we do include the
convolution with the beam.} We use the DSHARP opacities, which vary strongly
over the grain size range $[a_{\mathrm{min}},a_{\mathrm{max}}]$ we take. 

\begin{figure}
\centerline{\includegraphics[width=0.5\textwidth]{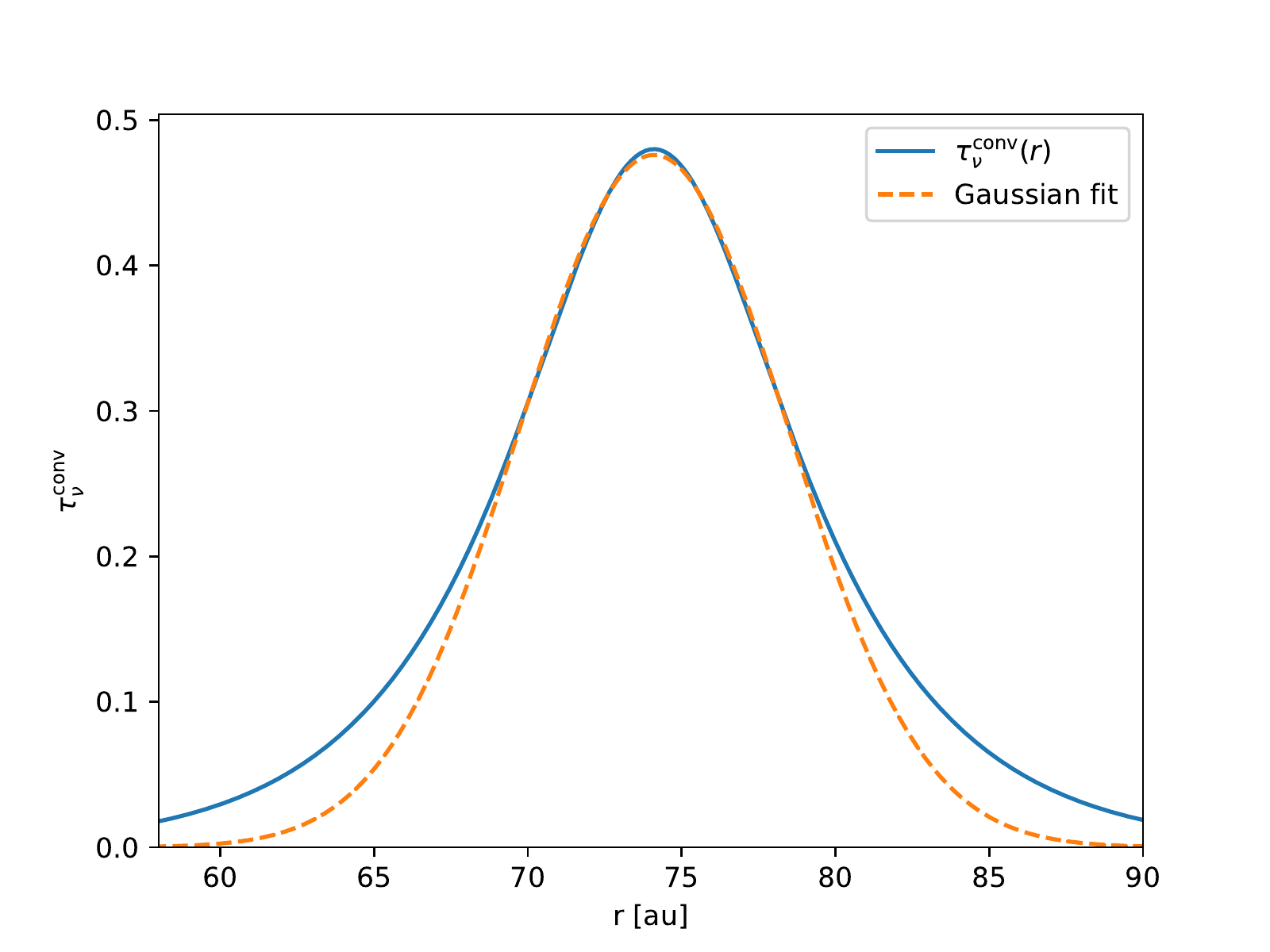}}
\caption{\label{fig-anmodel-distribution-total-tau}Total optical depth profile
  of the dust trapping model with a size distribution (solid line). The dashed
  line shows the Gauss curve that best fits the core of the profile.}
\end{figure}

The total optical depth profile
\revised{of this model is} shown in
Fig.~\ref{fig-anmodel-distribution-total-tau}. To see if this profile displays
excess emission in the wings, we fitted a Gaussian to the core of the profile,
in the same manner as we did in Section \ref{sec-gauss-fits}. We find indeed
that the core behaves nicely as a Gaussian, while the wings have excess, as
expected. The total required dust mass increases to 110 $M_{\oplus}$.
\removed{We see that a simple powerlaw dust distribution can explain excess emission in
the wings. The excess is of similar relative strength as the excesses seen in
ring 1 of AS 209, in the outer wing of ring 2 of the same source, the inner wing
of the Elias 24 ring (but not the outer plateau in that source), the outer wing
of the ring of GW Lup, and the inner/outer wing of ring 1/2 of HD 143006. In
contrast, the excess in the outer wing of ring 2 of HD 163296 is substantially
stronger. Some experimentation with the model shows that such a strong excess
requires the powerlaw to go all the way down to grain sizes of 1 $\mu$m, which
are the grains that are also seen in optical/near-infrared images.}

\revised{However, the model is symmetric, so it cannot explain the
  asymmetric excess of most rings. Some rings even show excess only on one
  side. In Section \ref{sec-planet-gap} we will address another scenario
  for the excess emission, which can explain also the asymmetry.}

\begin{figure}
\centerline{\includegraphics[width=0.47\textwidth]{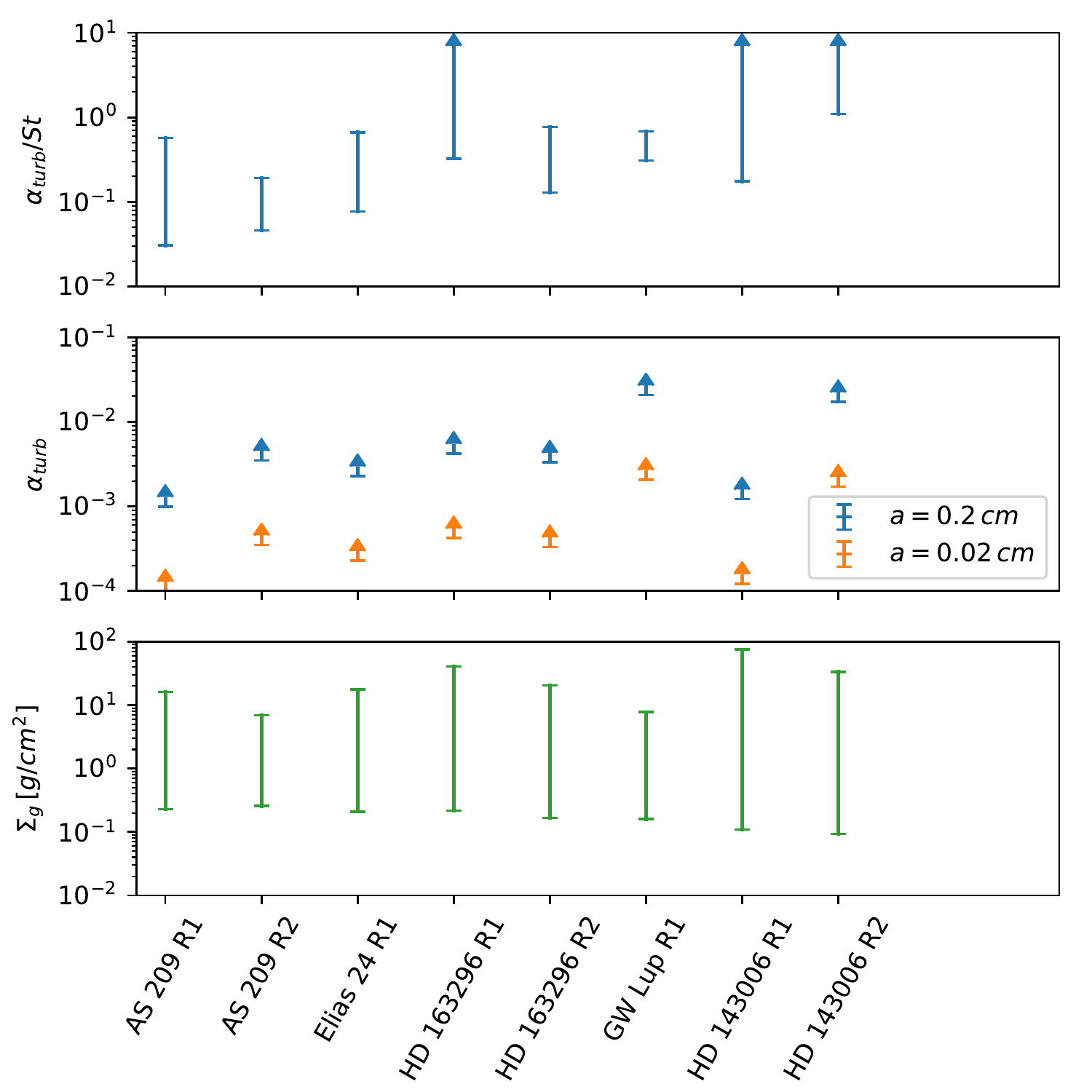}}
\caption{\label{fig-summary-2}Summary of the numbers resulting from the Gaussian
  dust trapping model analysis, as listed in Table
  \ref{tab-ring-model-limits}. Top: The values of
  $\alpha_{\mathrm{turb}}/\mathrm{St}$ found for the rings. Middle: The values
  of $\alpha_{\mathrm{turb}}/a_{\mathrm{grain}}$ for two choices of grain size,
  for the choice of $w=w_{\mathrm{max}}$ (leading to lowest possible values of
  $\alpha_{\mathrm{turb}}/a_{\mathrm{grain}}$). Bottom: Inferred range of gas
  surface density (bottom limit: $\Sigma_g\ge \Sigma_d$, top limit:
  gravitational instability).}
\end{figure}

The most important results we obtained in this section are summarized in
Fig.~\ref{fig-summary-2}.

\section{Planet gaps as dust traps and deviations from Gaussian shape}
\label{sec-planet-gap}
So far our models of dust trapping were quite idealized, in particular the
assumption of a gaussian pressure bump. In reality the radial pressure profile
is presumably better described by a smooth background profile with perturbations
imposed on it. The background profile could be, for instance, a powerlaw like
$p(r)\propto r^{-k}$ with index $k$ being $k=-2.5$. The perturbation could then
be a pressure bump or a pressure dip, \revised{the latter being the case for
  a planetary gap}. Given that the overall background pressure declines with
increasing $r$, such a dip/gap, if strong enough, could lead to a local pressure
maximum at the outer edge of the gap.  That would then be where the dust gets
trapped \citep[e.g.][]{2006MNRAS.373.1619R, 2012ApJ...755....6Z,
  2012A&A...545A..81P}. This pressure maximum would then not be symmetric like
the Gaussian pressure bump model of Section \ref{sec-rings-as-dust-traps}, but
instead is likely to be shallower on the outside and steeper on the inside.

As we know from the analysis of Section \ref{sec-rings-as-dust-traps}, the
widths of the dust rings of our sample are, in most cases, not very much
narrower than the widths of the gas pressure bumps (see $w_d/w$ column in Table
\ref{tab-ring-model-limits}). That means that the deviation of the gas pressure
bump from a Gaussian profile may affect the shape of the dust ring profile
too. If $w_d/w$ \revised{were to be} very small, the dust is only sensitive to the very peak of
the pressure bump profile. The larger $w_d/w$ is, the more the dust ``feels''
any non-Gaussian deviations in the wings of the bump. For AS 209, with $w_d/w$
\revised{of the order of} 0.2 for both rings \revised{(for the choice $w=w_{\mathrm{max}}$,
  see Table \ref{tab-ring-model-limits}, column 10)},
we thus expect the dust ring profiles to be
closer to Gaussian shape (modulo grain size distribution effects) than for HD
163296, for example.

\removed{This means that we may want to investigate if the slight deviations from
Gaussian profile seen in our dust ring sample (mainly the excess emission in the
wings) could be used to learn about the shape of the gas pressure bump.  If
successful, this may even give indications as to the origin of the pressure
bump. For instance: could the we find evidence for the bumps to be caused by a
planet opening up a gap?}

\revised{The question is: what do the wing-excesses in our ring sample
  tell us about the shape of the underlying pressure bump? And can we learn
  about its origin?}

Rather than addressing this question in a very general manner, we will start
straight from the scenario of a gap-opening planet. In another paper of this
series \citep{dsharp:zhang}, this hypothesis is investigated with detailed
hydrodynamic simulations of planet-disk interaction. Here, instead, we will
reduce this hypothesis to a very rudimentary model: a Gaussian dip in an
otherwise smoothly declining pressure profile. This produces an asymmetric
pressure bump at the outer edge of the gap.

\revised{The problem is now no longer a local one, but a global one: all the
  dust beyond the gap may, in time, drift into the dust trap and add to its
  mass.} We are forced to leave analytical modeling behind and employ numerical
techniques.

Our model is a 1-D viscous disk evolution model with a single dust component
added, which can radially drift and will be prone to radial turbulent
mixing. The equations of this model are standard, and have been repeated
numerous times in the literature \citep[e.g.][]{1976PThPh..56.1756A,
  2007A&A...469.1169B, 2007ApJ...671.2091G, 2010A&A...513A..79B,
  2012ApJ...755....6Z, 2016A&A...589A..15S}. Here we repeat the basic ones. The
gas surface density obeys
\begin{equation}
  \frac{\partial\Sigma_g}{\partial t} +
  \frac{1}{r}\frac{\partial\left(r\,\Sigma_g\,v_{gr}\right)}{\partial r} = 0
\end{equation}
with the radial gas velocity $v_{gr}$ given by
\begin{equation}\label{eq-radial-gas-velocity}
v_{gr}=-\frac{3}{\Sigma_g\sqrt{r}}\frac{\partial(\sqrt{r}\,\Sigma_g\nu_{\mathrm{turb}})}{\partial r}
\end{equation}
with $\nu_{\mathrm{turb}}=\alpha_{\mathrm{turb}} c_s^2/\Omega_K$ the turbulent viscosity of the disk.
The dust surface density obeys
\begin{equation}
  \frac{\partial\Sigma_d}{\partial t} +
  \frac{1}{r}\frac{\partial\left(r\,\Sigma_d\,v_{dr}\right)}{\partial r} =
  \frac{1}{r}\frac{\partial}{\partial r}\left[rD\Sigma_g\frac{\partial}{\partial r}
    \left(\frac{\Sigma_d}{\Sigma_g}\right)\right]
\end{equation}
with the radial dust velocity $v_{dr}$ given by
\begin{equation}\label{eq-radial-dust-velocity}
  v_{dr}=\frac{1}{1+\mathrm{St}^2}\,v_{gr}+\frac{1}{\mathrm{St}^{-1}+\mathrm{St}}
  \frac{c_s^2}{\Omega_Kr}\frac{d\ln p}{d\ln r}
\end{equation}
and the turbulent diffusion constant $D=\nu_{\mathrm{turb}}/(1+\mathrm{St}^2)$.
\removed{All this was
implemented in a new publicly available Python tool set called DISKLAB, the
details of which will be described in a separate paper
\citep{disklab:dullemondbirnstiel}.}

We will show here only a single example model, applied to ring 2 of HD
163296. An extensive study, applied to all the rings of this sample, will be
presented in a forthcoming paper. For our example model we set up a disk
according to the classic Lynden-Bell \& Pringle model
\citep{1974MNRAS.168..603L, 1998ApJ...495..385H} with an initial radius of 100
au, an initial disk mass of $10^{-1}\,M_\odot$. The temperature profile follows
the flaring angle recipe (Eq.~\ref{eq-disk-temperature-model}) with
$\varphi=0.02$ at all times, and the turbulence parameter is set to
$\alpha_{\mathrm{turb}}=10^{-2}$. We make a Gaussian dent into the disk model
at $r_p=85\,\mathrm{au}$ by defining a factor $F(r)$ 
\begin{equation}
F(r) = \exp\left[-f\exp\left(-\frac{(r-r_p)^2}{2w_\mathrm{gap}^2}\right)\right]
\end{equation}
such that
\begin{equation}
\Sigma_g(r) = \Sigma_{g0}(r)\, F(r)
\end{equation}
where $\Sigma_{g0}(r)$ is the unperturbed disk. We take the width of the gap to
be $w_{\mathrm{gap}}=6\;\mathrm{au}$ and the depth to be $f=2$. If we would viscously
evolve the disk without accounting for the continuous gap-opening force by the
planet, this initial gap would quickly be closed. To keep the gap open, without
having to include the complexities of planet-disk interaction (which anyway would
require at least a 2-D analysis), we apply the trick to replace the disk viscosity
(but not the turbulent mixing parameter) with
\begin{equation}
\nu_{\mathrm{turb}}(r) = \nu_{\mathrm{turb},0}(r) / F(r)
\end{equation}
Now we add the dust with an initial dust-to-gas ratio of $1:100$. As a grain
size we take $a_{\mathrm{grain}}=4\times 10^{-3}\,\mathrm{cm}$. We do not include
grain growth in this model.

\begin{figure}
\centerline{\includegraphics[width=0.5\textwidth]{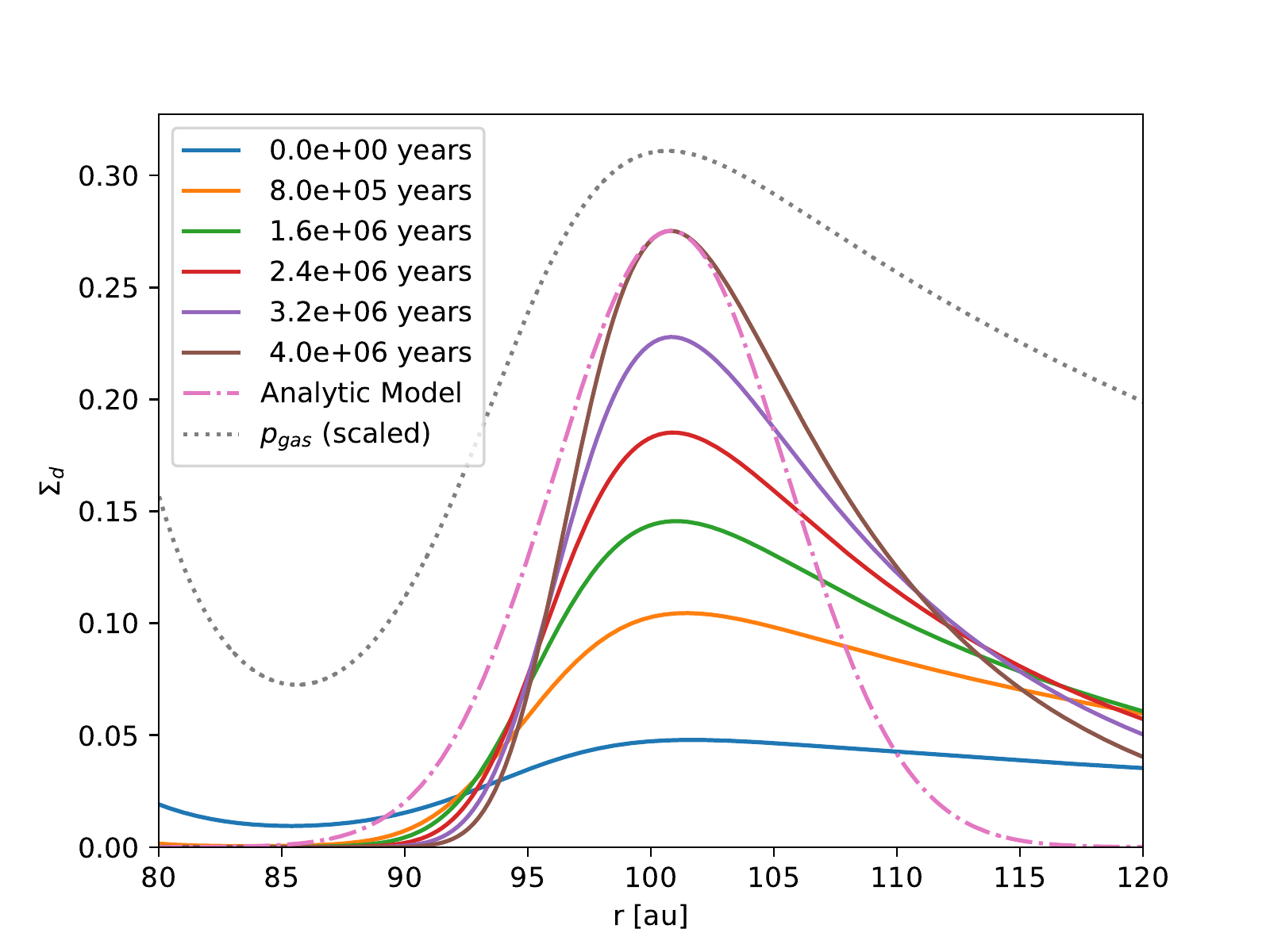}}
\caption{\label{fig-1d-drift-model-hd163296-ring2}\newstuff{Result of the numerical dust
  drift model described in Section \ref{sec-planet-gap}. Shown is the dust
  surface density of the dust near ring 2 of HD 163296 as it piles up in the
  pressure bump induced by the gap centered at 85 au. The dotted line shows, in
  a rescaled manner, the midplane gas pressure profile. The dot-dashed line
  shows the analytic solution of Section \ref{sec-rings-as-dust-traps},
  normalized to the final curve of the numerical model.}}
\end{figure}

The results of this model are shown in
  Fig.~\ref{fig-1d-drift-model-hd163296-ring2}. One can see that, as expected,
  the dust drifts into the local pressure peak located at $r_0=
  101\,\mathrm{au}$. As time goes by, more and more dust piles up there. The
  dust trap essentially collects all the dust from the outer disk regions. At 4
  Myr the dust pile-up is still on-going and no steady state is reached
  yet. This is due to our choice of relatively small dust grains. Had we chosen
  larger ones, the shape would have more quickly found its equilibrium shape,
  but it would have been significantly narrower, which is inconsistent with the
  observed dust ring width of ring 2 of HD 163296.

Overplotted is the analytic solution of Section
  \ref{sec-rings-as-dust-traps}. This solution needs a value of $w$, which we
  numerically compute from the second derivative of the midplane pressure
  profile: $w=\sqrt{-p(r_0)/(d^2p(r)/dr^2)_{r=r_0}}$. We see that the width of
  the numerical profile of the dust surface density is more or less consistent
  with the analytic result, but its shape is much steeper inside of $r_0$, and
  much shallower outside. This has two causes. One cause is the fact that the
  pressure profile is not a Gaussian, but is asymmetric. The other is that even
  at 4 Myr there is still dust flowing into the dust trap, in particular from
  the outside. The continuing steepening on both sides shows that the influx of
  dust declines with time, as the dust inside and outside of the bump gets
  depleted.

This section shows the limitations of the analytic solutions of
  Section \ref{sec-rings-as-dust-traps}. While the overall derived quantities
  such as the width of the dust ring are fairly well described by the
  analytic model, the deviations from Gaussian shape may not only be a result
  of the grain size distribution, but, as we see in this Section, also due to
  the non-Gaussian shape of the pressure bump and to the fact that the dust
  has not yet reached an equilibrium state.

\removed{The reasons for this can be traced back to the inferred lower limit
  values of $w_d/w$ from Section \ref{sec-rings-as-dust-traps} (see Table
  \ref{tab-ring-model-limits}), or equivalently the lower limit set on the ratio
  of $\alpha_{\mathrm{turb}}/a_{\mathrm{grain}}$ through
  Eq.~(\ref{eq-limit-on-turbulence}). For many rings, these values are not very
  small compared to 1. If we would put large grains into our model, for instance
  centimeter size grains, then $w_d/w$ would be much smaller, and the profile
  would be nearly Gaussian and steady state, even if the gas pressure bump is
  not Gaussian. However, such small $w_d/w$ values are excluded from the
  observations (the smallest one being 0.17 for ring 1 of AS 209).}

\removed{In other words: the combination of large grains $a_{\mathrm{grain}}\gg
  10^{-1}\,\mathrm{cm}$ and low turbulence $\alpha_{\mathrm{turb}}\ll 5\times
  10^{-4}$ can be excluded by the data (see Section
  \ref{sec-application-to-rings}). For instance, for HD 163296 ring 2 this lower
  limit is $\alpha/\mathrm{St}\gtrsim 0.13$. If $\alpha$ were to be $10^{-4}$,
  then $\mathrm{St}$ must be $\sim 7\times 10^{-4}$, which is quite low. With
  Eq.~(\ref{eq-agrain-to-stokes}) and $\Sigma_g$ given by the highest estimate
  from Table \ref{tab-ring-model-limits} (i.e.~$\Sigma_g\simeq
  20\,\mathrm{g/cm}^2$), we find that this Stokes number is consistent with
  grains up to, but not larger than, $a_{\mathrm{grain}}\lesssim 5\times
  10^{-3}\,\mathrm{cm}$.}

Given that the complexity of the numeric model is much higher than our analytic
models, we defer a more detailed parameter study and application of this model
to the DSHARP sample to a follow-up paper.

\section{Discussion}

\subsection{Why are most rings so ``fine-tuned''?}
It is rather striking that in the analysis of the rings up to this point we have
found several rather ``fine-tuned'' properties. For instance, the rings in the
disks around AS 209, Elias 24, GW Lup and the inner ring in the disk around HD
143006 have a width that is only roughly twice the beam size (between $1.6$ and
$2.7$ times, to be precise), but none are unresolved. Given the small sample,
and the fact that we selected isolated rings, it is very well possible that this
is just coincidence. The fact that some rings (in particular the inner ring of
HD 163296) are clearly much wider, lends some support to this.

The derived peak optical depths for most sources (except HD 143006), assuming
our model of the dust temperature is correct, hover around 0.4, i.e.\ just in
between the optically thin and optically thick regime. \revised{This also
  appears rather fine-tuned. Part of the explanation could be the fact that we
  selected the strongest-contrast rings in the DSHARP sample for our analysis.
  That may explain why none of our rings have very low optical depth. But it
  does not explain why none of them are very optically thick (flat-topped).}

Finally, many of the ring profiles are remarkably similar to a Gaussian shape.
\revised{This may be due to the fact that the rings are only a few beams
  wide, which may make non-Gaussian profiles appear more Gaussian after
  convolution. But it is unclear whether this explanation is sufficient.}

\revised{We therefore conclude that we do not know for sure whether the
  ``fine-tunedness'' of the rings in our sample is a real signal with a physical
  meaning, or an artifact of some kind. The question is, to which extent this
uncertainty could affect our conclusions.}

\revised{One of the main conclusions of our study is}
the fact that all the rings in our subsample are spatially resolved,
\revised{which} shows that
the dust trapping is not effective enough to produce very thin dust rings with
$w_d\ll h_p$. This is an important conclusion, which is also reflected in the
typical values of $\alpha_{\mathrm{turb}}/\mathrm{St}$ we derived (Table
\ref{tab-ring-model-limits}). \revised{Fortunately, this conclusion does
  not rely only on the measurement of $w_d$. It} is also supported by a flux
argument: The intensity before convolution cannot exceed the Planck function. So
assuming that our temperature estimate is correct, the minimal full width
$\Delta r$ of the dust ring would then be $\Delta r=\sqrt{2\pi}\sigma
A/B_\nu(T_d)$, where the factor $\sqrt{2\pi}$ originates from the integral over
the Gauss curve. For ring 1 of AS 209, for instance, this gives a width $\Delta
r\simeq 3.36\,\mathrm{au}$. This is about half the \fwhm{} of the current Gauss
estimate. In other words: even if, hypothetically, our measurement of the width
of the rings is entirely wrong, the fact that the rings are so bright (only
about a factor of 2 below the Planck function) shows that the rings cannot be
much narrower than the beam.

\subsection{Can a resolved ring be in fact a blend of several unresolved rings?}
\removed{Measuring the width of rings with finite-resolution observations is, of
  course, limited to ring widths that are larger than the ``beam''. However,
  even for cases where the ring width clearly exceeds the beam size by a factor
  of a few or more, one may ask the question:} \revised{Are the rings we see
  truly single rings, or could they also be made up of a concentric series of
radially} unresolved rings that are blended into a single ring due to the beam
convolution? It is, of course, hard to answer in general, since we have no
observational means to resolve structures of sub-beam size.

But from the perspective of particle trapping by a pressure bump this question
can be rigorously answered. A long-lived radial pressure perturbation in a
protoplanetary disk cannot be much narrower than about a pressure scale height
$h_p(r)$ \citep{2016ApJ...823...84O}. A dust ring produced by dust trapping in
this pressure bump may become rather narrow, dependent on a variety of
parameters, as discussed in Section
\ref{sec-steady-state-analytic-trap-model}. But there can not be more than a
single such dust ring in each pressure bump.

For the wide rings of HD 163296, even under the most optimistically low disk
temperature (e.g.~10 K) the pressure scale height at rings 1 and 2 are 2.4 au
and 4.3 au, respectively, which correspond to \fwhm{} widths of 55 and 101
milliarcseconds, respectively\removed{ (where we have multiplied by 2.355 to obtain the
full-width at half-maximum corresponding to a gaussian with $h_p$ as standard
deviation width)}. Clearly the ALMA observations in band 6, with \fwhm{} beam
size of 51 mas, spatially resolve the pressure scale height. This means that
the ring separation will be spatially resolved by ALMA, ruling out the
possibility that the wide rings are made up of a multitude of narrow rings, at
least in the dust trapping scenario.

However, the rings may be made up of many unresolved clumps, such as those
produced by the streaming instability. Whether the
presence of such clumpy structure has observable consequences, in spite of
the clumps being spatially unresolved, is an issue that requires deeper
study. But one may speculate that the self-regulation mechanism of the
streaming instability, as discussed above, may also lead to a self-regulation
of the optical depth or, equivalently, the covering fraction of unresolved
optically thick clumps. 

\subsection{Condition for the streaming instability}
The ``streaming instability'' and related processes \citep{2005ApJ...620..459Y,
  2007ApJ...662..627J, 2010ApJ...722.1437B, 2013MNRAS.434.1460K,
  2017A&A...597A..69S, 2018ApJ...861...47S} play a fundamental role in the
theory of planet formation. Dust traps may be ideal places for this process to
operate, because in those regions one can expect the local dust-to-gas ratio to
be strongly enhanced over the background. There is the concern that at the
precise location of the pressure maximum the streaming instability is killed
because the gas orbits exactly \revised{with Kepler velocity} there. But
slightly adjacent to the pressure peak the deviation from Keplerian motion is
strong, and may drive such an instability. To keep dust in those adjacent
regions, turbulence is required to counteract the trapping. If this turbulence
is caused by the streaming instability itself, this is a bit of a
``chicken-or-egg'' issue. \citet{2018MNRAS.473..796A} report a linear stability
analysis that indicates that the streaming instability can occur in pressure
bumps. \citet{2015ApJ...804...35R} present simulations of particle trapping and
streaming instability in a vortex, which is in many ways similar to the dust
traps we study in this paper. But the final word on this matter has not yet been
said. Let us, for the purpose of the argument, assume that the streaming
instability, \revised{and the related process of gravoturbulent planetesimal
  formation \citep{2007Natur.448.1022J},} can indeed occur in a pressure bump.

In the literature it is often
mentioned that the streaming instability requires a dust-to-gas surface density
ratio of $\Sigma_{\mathrm{d}}/\Sigma_{\mathrm{g}}\gtrsim 0.02$ or higher to
operate \citep{2010ApJ...722.1437B}.  This can, however, not be directly
compared to our models, because this value of $0.02$ was found for models
without any pre-determined turbulence. The turbulence in those models was
induced by the streaming instability itself. In our analytic model, on the other
hand, we set the turbulence strength by hand, by setting
$\alpha_{\mathrm{turb}}$ to some value. In essence, we assume that there is
another source of turbulence, such as the magnetorotational instability or the
vertical shear instability, that determines the mixing of the dust in the disk
\citep[see e.g.][]{2018arXiv180808681L}.

According to \citet{2005ApJ...620..459Y} the true criterion for the onset of the
streaming instability is the ratio of dust and gas {\em volume} densities
$\rho_{\mathrm{d}}/\rho_{\mathrm{g}}\gtrsim 1$. The midplane volume density
ratio for a single grain species with midplane Stokes number $\mathrm{St}\ll 1$,
and given surface density ratio $\Sigma_{\mathrm{d}}/\Sigma_{\mathrm{g}}$,
depends on the turbulent strength as
\begin{equation}\label{eq-dtg-sig-vs-rho}
  \frac{\rho_{\mathrm{d}}}{\rho_{\mathrm{g}}}\simeq
  \left(1+\frac{\mathrm{St}}{\alpha_{\mathrm{turb}}}\right)^{1/2}
  \frac{\Sigma_{\mathrm{d}}}{\Sigma_{\mathrm{g}}}
\end{equation}
\afterintrev{(see Eq.~\ref{eq-hd-afo-hp-psi}, \revised{and setting $\mathrm{Sc}=1$}).}
The criterion of $\Sigma_{\mathrm{d}}/\Sigma_{\mathrm{g}}\gtrsim 0.02$ mentioned
in the literature thus relates to the criterion
$\rho_{\mathrm{d}}/\rho_{\mathrm{g}}\gtrsim 1$ via the turbulent strength and
the Stokes number. Given that we do not compute the turbulent strength,
but prescribe it, we should rely on the more fundamental volume density
criterion of \citet{2005ApJ...620..459Y} to assess whether the dust in our
model triggers the streaming instability or not.

To get some numbers, let us take ring 1 of AS 209. Let us assume the widest
possible pressure bump, i.e.\ $w=w_{\mathrm{max}}$, for which the ratios
$\alpha_{\mathrm{turb}}/\mathrm{St}=3.1\times 10^{-2}$, as listed in Table
\ref{tab-ring-model-limits}. This leads, with Eq.~(\ref{eq-dtg-sig-vs-rho}), to
a dust-to-gas volume density ratio that is 5.8 times larger than the dust-to-gas
surface density ratio. This means that the criterion by
\citet{2005ApJ...620..459Y} is triggered if
$\Sigma_{\mathrm{d}}/\Sigma_{\mathrm{g}}\gtrsim 0.17$. Given that
$\Sigma_{\mathrm{d}}=\Sigma_{\mathrm{g,min}}$ (by definition of the latter), we
can look up its value in Table \ref{tab-ring-model-limits} and find that for
$\Sigma_{\mathrm{g}}\lesssim 1.4\,\mathrm{g}\,\mathrm{cm}^{-2}$ the streaming
instability will be triggered. Given that the disk becomes gravitationally
unstable for $\Sigma_{\mathrm{g}}\gtrsim 16 \,\mathrm{g}\,\mathrm{cm}^{-2}$,
this leaves only little more than a factor of 10 room for $\Sigma_{\mathrm{g}}$
to avoid either the streaming instability or the gravitational instability.
\revised{Note that if we take a narrower pressure bump
  (e.g.~$w=w_{\mathrm{min}}$), the ratio $\alpha_{\mathrm{turb}}/\mathrm{St}$
  increases, making it harder for the streaming instability to set in.}

In the end we cannot, therefore, say with any certainty whether the streaming
instability is operating in these rings or not. But we do find that the
likelihood that the conditions are triggered are realistic. The rings we see may
therefore consist of unresolved clumps, in which planetesimals may form
\citep{2007Natur.448.1022J}. 

However, one may then wonder why this does not
immediately convert all dust into planetesimals. This may be due to a
self-regulation effect: once a certain fraction of the dust is converted into
planetesimals, the remaining dust is no longer dense enough to trigger strong
enough clumping \citep{2014A&A...572A..78D}.

\subsection{Caveats of the models}
\label{sec-caveats}
\newstuff{This paper is meant as the initial step of a bottom-up investigation
  of the ringlike structures found in the DSHARP campaign: starting with the
  simplest analytic estimates, and building up the complexity and realism of the
  models, so that is becomes clearer what the data tell us -- and what not.}

\newstuff{Among the important aspects we have not treated in this paper are: the
  dust back-reaction onto the gas \citep[e.g.][]{2007ApJ...662..627J,
    2017MNRAS.467.1984G, 2017ApJ...844..142K}, the origin of the pressure bumps
  and/or gaps \citep[e.g.][]{2012A&A...545A..81P, 2016A&A...589A..87B,
    2016AJ....152..184T, 2018A&A...609A..50D}, the detailed shape of planetary
  gaps \citep[e.g.][]{2017PASJ...69...97K, dsharp:zhang}, 2-D and 3-D effects,
  full radiative transfer \citep[e.g.][]{2013A&A...549A.124B,
    2013A&A...560A..43F}, dust growth and fragmentation
  \citep[e.g.][]{2010A&A...513A..79B, 2012ApJ...752..106O}, and many other
  things.}

\newstuff{Also, if we would include, for the analytic models of the dust traps,
  a temperature gradient and a background density gradient, the results may be
  affected. In particular the exact location of the pressure peak may shift.}

\newstuff{This paper is therefore not meant to give definitive numbers or
  conclusions. Rather, it is meant as a starting point of more complex
  modeling campaigns. One such more complex modeling campaign is the
  hydrodynamic planet-disk interaction paper by \citet{dsharp:zhang}.}

\section{Conclusions}
\afterintrev{We studied the radial structure of the eight most prominent
  dust rings from the DSHARP sample, and investigated to which extent they are
  consistent with, and/or indications of, being dust traps.}

We can summarize our conclusions as follows:
\begin{enumerate}
\item For the rings in AS 209, Elias 24, the outer ring of HD 163296, and the
  ring of GW Lup the width is narrower than the estimated pressure scale
  height. This is strong evidence for dust trapping being at work.
\item \afterintrev{For none of the 8 rings studied in this paper we found
  evidence against dust trapping.}
\item The dust trapping may explain their longevity, given the fact that dust
  grains tend to drift into the star on a short time scale in the absense of
  dust traps \citep{2012A&A...538A.114P}.
\item All rings are radially resolved, \afterintrev{by factors $\sigma/\sigma_b$
  ranging from 1.6 (ring B41 of HD 143006) to 3.2 (ring B67 of HD 163296).
  When comparing the implied width of the dust ring $w_d$ to the largest
  plausible width of the gas pressure bump $w$, we find that the strongest dust
  trapping occurs in AS 209, with $w_d/w$ ratios of 0.17 and \revised{0.21} for rings 1
  and 2, respectively. For the other rings we find larger $w_d/w$ values. This
  indicates that turbulent mixing is at play, preventing the dust from being
  compressed into an even narrower ring. Or it could mean that the dust grains
  are so small, that they have not yet reached drift-mixing equilibrium.}
\item All rings have absorption optical depths in the range 0.2 to 0.5. When
  scattering is included, the total optical depth may even be higher. But we can
  exclude complete saturation: none of the rings are completely optically
  thick. But until we have spectral information we cannot exclude
    the rings to consist of unresolved optically thick clumps with a beam
    filling factor in the range 0.2 to 0.5.
\item The narrow range in optical depth suggests that some sort of
  self-regulation mechanism is operating, perhaps related to planet formation
  processes.
\item The radial shape of the dust emission rings can mostly be described by a
  Gaussian \revised{profile}, consistent with dust trapping of a single grain size in a
  Gaussian pressure bump, in which the trapping force is in equilibrium with
  turbulent spreading.  In the wings some profiles have excess emission, which
  may be an indication of a grain size distribution, with small grains being
  spread out wider than the big ones. However, the excess is more often seen on
  the outside than on the inside in the rings in our sample.  This may be an
  indication of ongoing influx of dust from larger radii into the dust trap. Our
  simple numerical model of dust trapping in the outer edge pressure bump of a
  planetary gap also indicates that the asymmetry of the gas pressure bump,
  being steeper on the inside than the outside, may be reflected in the dust as
  well.
\item The dust masses stored in the rings are of the order of tens of Earth
  masses. The gas surface density is limited from below by the demand that
  it should be at least larger than the dust surface density. From above it
  is limited by the gravitational stability criterion. This leaves a range
  of two orders of magnitude for the gas surface density.
\item The high dust mass trapped in these rings makes it plausible
  that the conditions for the streaming instability are met (if the streaming
  instability indeed works in a pressure trap). This could perhaps be the
  source of turbulence that prevents the dust ring from becoming ultra-narrow.
\item We estimate a lower limit of $\alpha_{\mathrm{turb}}\simeq
  10^{-4}$, but much larger values of $\alpha_{\mathrm{turb}}$ are also consistent with our data.
  We need spectral information to constrain the grain size and dynamic
  information to constrain the width of the gas pressure bump.
\item\label{concl-alphastokes} Given the not so small values of $w_d/w$ inferred for most
  rings, the combination of very low $\alpha_{\mathrm{turb}}\ll 5\times 10^{-4}$
  and very large grains $a_{\mathrm{grain}}\gg 0.1\,\mathrm{cm}$ can be excluded
  by the data. To be more precise, we can exclude
  \revised{$\alpha_{\mathrm{turb}}\lesssim (a_{\mathrm{grain}}/0.02\,\mathrm{cm})\,\alpha_{\mathrm{exmp}}$,}
  with $\alpha_{\mathrm{exmp}}$ given in Table \ref{tab-ring-model-limits}.
\item \revised{In addition to the dynamical arguments from conclusion
  \ref{concl-alphastokes}, from opacity arguments we can put strong upper limits
  on the grain size of 1 cm to half a meter, depending on the ring.}
\item Our analysis does not generate conclusions as to the origin
  of the gas pressure maxima which trap the dust. However, our scenario is
  completely consistent with their origin being the formation of a planetary
  gap. If the unperturbed disk has $dp/dr<0$, then a planetary gap would produce
  a pressure bump at the outer edge of the gap. See \citet{dsharp:zhang} for a
  detailed discussion of this scenario in the context of the DSHARP survey.
\end{enumerate}

\begin{acknowledgements}
C.P.D.\ acknowledges support by the German Science Foundation (DFG) Research
Unit FOR 2634, grants DU 414/22-1 and DU 414/23-1. 
T.B.\ acknowledges funding from the European Research Council (ERC) under the
European Union's Horizon 2020 research and innovation programme under grant
agreement No 714769. S.A. and J.H. acknowledge funding support from the
National Aeronautics and Space Administration under grant No. 17-XRP17\_2-0012
issued through the Exoplanets Research Program.
J.H. acknowledges support from the National Science Foundation Graduate Research
Fellowship under Grant No. DGE-1144152.
L.R. acknowledges support from the ngVLA Community Studies program, coordinated
by the National Radio Astronomy Observatory, which is a facility of the National
Science Foundation operated under cooperative agreement by Associated
Universities, Inc.
V.V.G. and J.C acknowledge support from the National Aeronautics and Space
Administration under grant No. 15XRP15\_20140 issued through the Exoplanets
Research Program.
Z.Z. and S.Z.\ acknowledge support from the National Aeronautics and Space
Administration through the Astrophysics Theory Program with Grant No. NNX17AK40G
and Sloan Research Fellowship. Simulations are carried out with the support from
the Texas Advanced Computing Center (TACC) at The University of Texas at Austin
through XSEDE grant TG- AST130002.
M.B. acknowledges funding from ANR of France under contract number
ANR-16-CE31-0013 (Planet Forming disks).
L.P. acknowledges support from CONICYT project Basal AFB-17002 and from
FCFM/U. de Chile Fondo de Instalaci\'on Acad\'emica.
A.I. acknowledges support from the National Aeronautics and Space Administration
under grant No. NNX15AB06G issued through the Origins of Solar Systems program,
and from the National Science Foundation under grant No. AST-1715719.
This paper makes use of ALMA
data
\dataset[ADS/JAO.ALMA\#2016.1.00484.L]{https://almascience.nrao.edu/aq/?project\_code=2016.1.00484.L}
\end{acknowledgements}

\begingroup
\bibliography{ms}
\endgroup

\appendix

\section{Symbols}
\label{sec-symbols-table}
\revised{
Since this paper contains many equations and symbols, here we present a summary
table of the symbols used.}

\begin{deluxetable}{lll}
\tablecaption{Symbols and their meaning.\label{tab-symbols}}
\tablecolumns{3}
\tablewidth{0pt}
\tablehead{
\colhead{Symbol} &
\colhead{Meaning} &
\colhead{Eq.~of definition}\\
}
\startdata
$\nu$, $\lambda$         & Frequency and wavelength of the observation & $\lambda=c/\nu\simeq 0.125\,\mathrm{cm}$\\
$I_{\nu}^{\mathrm{gauss}}$, $I_{\nu}^{\mathrm{gauss,dec}}$  & Gaussian fit to intensity profile, and its deconvolved version & Eq.~(\ref{eq-gauss-lin-br-temp}, \ref{eq-gauss-lin-br-temp-deconv})  \\
$A$, $A_{\mathrm{dec}}$    & Amplitude $A$ of Gaussian fit and its deconvolved version $A_{\mathrm{dec}}$  & Eqs.~(\ref{eq-gauss-lin-br-temp}, \ref{eq-a-deconv})  \\
$r_0$                    & Radius of ring at pressure peak  &  Eq.~(\ref{eq-gauss-lin-br-temp}) \\
$\sigma$                 & Width (standard deviation) of radial intensity profile of ring in au  &  Eq.~(\ref{eq-gauss-lin-br-temp}) \\
$b_{\mathrm{fwhm,as}}$, $\sigma_b$ & Beam FWHM in arcsec, and its standard deviation in au  & $\sigma_b=d_{\mathrm{pc}}b_{\mathrm{fwhm,as}}/2.355$ \\
$d_{\mathrm{pc}}$          & Distance in parsec &   \\
$w_d$                     & Width of the dust ring in au &  Eqs.~(\ref{eq-simple-deconvolve-gauss}, \ref{eq-wd-afo-w-psi}) \\
$w$, $w_{\mathrm{min}}$, $w_{\mathrm{min}}$  & Width of the gas ring, and its lower and upper limits &  Section \ref{sec-ranges-of-params} \\
$T_{\mathrm{g}}$, $T_{\mathrm{d}}$ & Midplane temperature in gas and dust  & Eq.~(\ref{eq-disk-temperature-model})  \\
$c_s$                    & Isothermal sound speed & $c_s=\sqrt{k_BT_{\mathrm{g}}/\mu m_p}$ \\
$\Omega_K$               & Kepler frequency & $\Omega_K=\sqrt{GM_{*}/r^3}$ \\
$h_p$, $h_d$             & Pressure scale height of the gas, and vertical height of the dust layer  & Eq.~(\ref{eq-hp-afo-t}, \ref{eq-hd-afo-hp-psi})  \\
$k_B$, $m_p$, $G$        & Natural constants: Boltzmann constant, proton mass, gravitational constant  &   \\
$\Sigma_{\mathrm{d}}$, $\Sigma_{\mathrm{d}}^{\optthin{}}$, $\Sigma_{\mathrm{d}}^{\mathrm{gauss}}$    & Dust surface density, its optically thin estimate, and its Gaussian fit & Eq.~(\ref{eq-optthin-conversion-linbright-sigmad})  \\
$\Sigma_{\mathrm{g}}$, $\Sigma_{\mathrm{g,min}}$, $\Sigma_{\mathrm{g,max}}$    & Gas surface density, and its lower and upper limits & Eqs.~(\ref{eq-sigmag-lower-lim}, \ref{eq-sigmag-upper-lim})  \\
$\rho_d$, $\rho_g$       & Dust and gas volume density at the midplane &   \\
$M_d$, $M_d^{\optthin{}}$   & Dust mass in the ring, and its optically thin estimate & Eqs.~(\ref{eq-dust-mass-estimate}, \ref{eq-dust-mass-estimate-real})\\
$B_\nu$                   & Planck function  &   \\
$\kappa_\nu^{\mathrm{abs}}$  & Dust absorption opacity  &   \\
$a$, $a_{\mathrm{min}}$, $a_{\mathrm{max}}$     & Dust grain radius, and its limits (for size distribution)  &   \\
$\tau_\nu^{\mathrm{peak}}$   & Optical depth at the peak of the ring  & Eq.~(\ref{eq-tau-estimate})  \\
$\tau_\nu(r)$             & Optical depth profile of the ring  & Eq.~(\ref{eq-tau-profile})  \\
$Q_{\mathrm{Toomre}}$      & Toomre parameter & Eq.~(\ref{eq-sigmag-upper-lim})\\
$\alpha_{\mathrm{turb}}$, $\alpha_{\mathrm{exmp}}$    & The turbulence $\alpha$-parameter, and its value for $a=0.02\,\mathrm{cm}$  &  Eq.~(\ref{eq-d-afo-alpha}) \\
$\mathrm{St}$            & Stokes number of the dust particles  &   \\
$\mathrm{Sc}$            & Schmidt number of the turbulence (usually set to 1)  &   \\
$\psi$                   & If $\psi>>1$: constant dust/gas ratio; if $\psi<<1$: strong dust trapping  & Eq.~(\ref{eq-psi-afo-alpha-sc-st})  \\
$\xi_{\mathrm{dust}}$      & Material density of the dust grains  &   \\
$v_{\mathrm{gr}}$, $v_{\mathrm{dr}}$     & Radial velocity of gas and dust, respectively  & Eqs.~(\ref{eq-radial-gas-velocity}, \ref{eq-radial-dust-velocity}) \\
$p$                     & Gas pressure at the midplane   &   \\
$\nu_{\mathrm{turb}}$    & Turbulent viscosity coefficient  & $\nu_{\mathrm{turb}}=\alpha_{\mathrm{turb}} c_s^2/\Omega_K$ \\
$D$                   & Turbulent diffusion coefficient  & $D=\nu_{\mathrm{turb}}/(1+\mathrm{St}^2)$  \\
$w_{\mathrm{gap}}$     & Width of the gap carved out by a planet  &   \\
\enddata
\end{deluxetable}

\section{Gauss fitting procedure}
\label{sec-gauss-fitting-procedure}
\removed{In this paper we study each dust ring individually, and try to understand it in
terms of the trapping of dust in a pressure trap. In Appendix
\ref{sec-steady-state-analytic-trap-model} we find that for a Gaussian pressure
bump the solution to the radial dust mixing and drift problem is, to first
approximation, also a Gaussian, albeit a narrower one. Fitting the dust trapping
model to the data can therefore be done in two stages: first fitting a Gaussian
radial profile to the observed rings, then interpreting these Gaussian fits
through the dust trapping model. In this Appendix we detail the procedure we
used to fit the radial intensity profiles of the rings with Gaussian profiles.}
The radial intensity profiles were
extracted from the images using a procedure similar to that described by
\citet{dsharp:huangrings}. This procedure involves the fitting of an ellipse to
describe the inclined ring shape, the deprojection into a \afterintrev{circular} ring, and the
averaging of the intensity along the ring. This averaging procedure enhances the
signal-to-noise ratio considerably, by a factor $\sqrt{N}$, where $N$ is the
number of beams that fit along the ring. We estimate the intrinsic noise simply
by computing the standard deviation along the ring. The resulting averaged
radial intensity profile $I_\nu(r)$ thus obtains also an error
estimate $\varepsilon(r)$, which is typically of the order of $\sim$1\% of
the peak intensity.

The rings display themselves as bumps in $I_\nu(r)$. We choose by eye
a radial domain around the bump where we believe a Gaussian description is justified.
The inner and outer radii of this domain are listed in Table \ref{tab-gauss-params}.
By choosing this domain we can select a specific ring to fit, which is not possible
when doing the fitting procedure in the uv-plane.

We now fit a Gaussian profile to this bump
\begin{equation}
I_\nu^{\mathrm{gauss}}(r) = A\exp\left(-\frac{(r-r_0)^2}{2\sigma^2}\right)
\end{equation}
We use the code {\small\sf emcee} \citep{2013PASP..125..306F} to perform a
Markov Chain Monte Carlo (MCMC) procedure to find the set of parameters
$(A,r_0,\sigma)$ which have the highest likelihood. The sampling of
$I_\nu^{\mathrm{gauss}}(r)$ is about $N\simeq 70$ points per beam
\afterintrev{in radial direction}.  \afterintrev{But of course these data points
  are not independent: there is only one independent measurement per beam (the
  multiple beams along each ring are already accounted for by the accordingly
  reduced error). We therefore have to multiply the error estimate of the
  datapoints by $\sqrt{70}$ before} feeding it into {\small\sf emcee}.

We use 100 walkers with 500 steps, and use the last 250 steps for our
statistics. The most likely parameter values and their error estimates are given in
Table \ref{tab-gauss-fit-errors}.

\begin{deluxetable*}{ccccc}[b!]
\tablecaption{The Gaussian fit values with their error
  estimates.\label{tab-gauss-fit-errors}}
\tablecolumns{5}
\tablewidth{0pt}
\tablehead{
\colhead{Source} &
\colhead{Ring} &
\colhead{$A$} &
\colhead{$r_0$} &
\colhead{$\sigma$} \\
}
\startdata
AS 209     & 1 & $  0.141_{-0.002}^{+0.002}$ & $ 74.180_{-0.074}^{+0.072}$ & $  3.976_{-0.108}^{+0.119}$  \\
AS 209     & 2 & $  0.114_{-0.001}^{+0.001}$ & $120.429_{-0.076}^{+0.078}$ & $  4.616_{-0.120}^{+0.131}$  \\
Elias 24   & 1 & $  0.228_{-0.002}^{+0.002}$ & $ 76.654_{-0.072}^{+0.068}$ & $  4.927_{-0.134}^{+0.133}$  \\
HD 163296  & 1 & $  0.358_{-0.003}^{+0.003}$ & $ 67.741_{-0.055}^{+0.056}$ & $  7.185_{-0.057}^{+0.061}$  \\
HD 163296  & 2 & $  0.215_{-0.002}^{+0.002}$ & $ 99.962_{-0.065}^{+0.068}$ & $  5.169_{-0.117}^{+0.126}$  \\
GW Lup     & 1 & $  0.054_{-0.001}^{+0.002}$ & $ 85.552_{-0.315}^{+0.443}$ & $  5.810_{-0.484}^{+0.672}$  \\
HD 143006  & 1 & $  0.138_{-0.003}^{+0.003}$ & $ 40.993_{-0.191}^{+0.238}$ & $  5.092_{-0.318}^{+0.403}$  \\
HD 143006  & 2 & $  0.107_{-0.002}^{+0.002}$ & $ 65.161_{-0.261}^{+0.247}$ & $  8.006_{-0.539}^{+0.676}$  \\
\enddata
\tablecomments{Error estimates are obtained from the MCMC procedure described in Appendix \ref{sec-gauss-fitting-procedure}.}
\end{deluxetable*}

\section{Comments on the Gaussian fitting in the image plane \revised{vs.~the uv-plane}}
\label{sec-comments-on-gauss}
For the interpretation of these rings in terms of dust trapping it is critical
to know the true width of the rings: whether they are radially resolved or
not. The ratio of the ring width in units of the effective beam size is listed
as $\sigma/\sigma_b$ listed in Table \ref{tab-gauss-params}.  This shows that
all rings are radially resolved, most of them by about 2$\ldots$3 beam
widths. Some rings are, however, only marginally resolved, such as ring 1 of HD
143006, which is only 1.6 beams wide. The closer $\sigma/\sigma_b$ is to 1, the
harder it is to derive the true width, because it requires an increasingly
precise understanding of the convolution kernel.

\revised{By comparing our inferred ring widths to those inferred in the uv
  plane, we can get an estimate of the reliability of our numbers. In the DSHARP
  series, three papers analyze rings from our subsample using model fitting in
  the uv-plane: \citet{dsharp:guzman} for AS 209, \citet{dsharp:isella} for HD
  163296, and \citet{dsharp:perez} for HD 143006.}

\revised{For AS 209 \citet{dsharp:guzman} derive a ring width that is 10\%
  narrower for ring 1 and 20\% narrower for ring 2 than in this paper.  For HD
  163296 \citet{dsharp:isella} find roughly the same width for ring 1, but a
  18\% wider ring 2. Finally, for HD 143006 \citet{dsharp:perez} find a 8\%
  wider ring 1, and a 30\% wider ring 2.}

\revised{For HD 143006, however, the rings are not very well separated,
meaning that the different fitting criteria between the method of this paper and
that of \citet{dsharp:perez} is likely responsible for the differences.}

It is \revised{clear} that the Gaussian fitting in this paper has its
limitations. First of all, it lies in the nature of fitting a Gaussian profile
to something non-Gaussian that there will be a region close to the peak where
the curve fits the Gaussian reasonably well, while the deviation will increase
the farther away from the peak one looks. This is particularly so
in the present case, since the fitting range was chosen to maximize the
similarity to the Gaussian shape near the peak. Secondly, we fit the Gaussians
in the image plane, not in the uv-plane. This means that we do not fit to the
actual data, but to a reconstruction of the data, which may add additional
sources of errors that are hard to identify.

\removed{For three sources of our sample,
however, the rings were also fitted in the uv plane, yielding compatible, though
not identical, results (\citet{dsharp:guzman} for AS 209, \citet{dsharp:isella}
for HD 163296; and \citet{dsharp:perez}).}

In Appendix \ref{sec-mock-ring-test} we show the results of a simple mock ring
test, showing that in principle the results derived from the data in the image
plane should be accurate enough for our purposes.

\section{Computing dust mass including mild optical depth effects}
\label{sec-compute-true-dust-mass}
\revised{Given that the shapes of the radial profiles are nearly Gaussian, we
  have been tempted to assume that the dust emission is optically thin, in which
  case Eq.~(\ref{eq-dust-mass-estimate}) gives the mass of dust in the ring
  $M_d^{\optthin{}}$. In reality the ring contains more mass, hidden by the
  optical depth effects. If we assume that the real dust radial profile is truly
  Gaussian (i.e.\ $\tau_\nu^{\mathrm{Gauss}}(r)$), this means that the putative
  Gaussian shape we observe is apparently not real. We see the function
  $(1-\exp(-\tau_\nu^{\mathrm{Gauss}}(r)))$ instead of
  $\tau_\nu^{\mathrm{Gauss}}(r)$. However, using numerical experimentation one
  can show that for mild optical depths, such a profile can be fitted reasonably
  well by an alternative Gaussian shape, with only minor deviations. This
  alternative Gaussian curve is slightly broader than
  $\tau_\nu^{\mathrm{Gauss}}(r)$ and has a substantially lower peak. For peak
  optical depths below unity the fit is remarkably good. We call this ``Gaussian
  mimicry'', because a non-Gaussian radial profile poses as a Gaussian.}

\revised{This means that we may think we are dealing with a Gaussian shape, but
  the Gaussian parameters (width and amplitude) are, in a manner of speaking,
  ``fake''. The peak of the real optical depth profile is, by definition,
  $\tau_\nu^{\mathrm{peak}}$. The peak of the mimicked Gaussian is
  approximately $(1-\exp(-\tau_\nu^{\mathrm{peak}}))$. If the width of the real
  Gaussian is $w_d^{\mathrm{true}}$, then the widths of the mimicked Gaussian
  has to be obtained through numerical calculation. We use the {\small\tt
    scipy.optimize.minimize()} function of the SciPy library of Python to fit a
  Gaussian to the $(1-\exp(-\tau_\nu^{\mathrm{peak}}))$ profile, which is the
  mimicked Gaussian. The numerically obtained widths $w_d^{\mathrm{mimick}}$ can be
  approximated by the following formula:
 \begin{equation}\label{eq-calw-approx}
   \frac{w^{\mathrm{mimick}}_d}{w^{\mathrm{true}}_d}\simeq {\cal W}\equiv \sqrt{2.15\,\ln\left(1+0.148\,\tau_\nu^{\mathrm{peak}}\right)+1}
 \end{equation}
 This ratio is typically between 1 and about 1.15. The Gaussian fitting of
 Section \ref{sec-gauss-fits} evidently yields $w_d^{\mathrm{mimick}}$. So using
 Eq.~(\ref{eq-calw-approx}) we can then compute from that $w_d^{\mathrm{true}}$.
 The optical-depth-corrected dust mass is then
\begin{equation}\label{eq-dust-mass-estimate-real}
  M_d^{\mathrm{true}} = M_d^{\optthin{}}\;\frac{1}{{\cal W}}\;
  \frac{\tau_\nu^{\mathrm{peak}}}{1-e^{-\tau_\nu^{\mathrm{peak}}}}
\end{equation}
where $M_d^{\optthin{}}$ is the optically thin mass estimate of
Eq.~(\ref{eq-dust-mass-estimate}). This optical-depth-corrected mass is also
listed in Table \ref{tab-gauss-params}. It is only up to 20\% higher than the
optically thin mass.}

These optical depth corrections are of course only valid if we assume
  a smooth distribution of dust. If the dust is distributed into a multitude of
  spatially unresolved optically thick clumps, then much more mass could
  conceivably be hidden in these clumps.

Note also that in dealing with the optical depth issues, we have so far only
concentrated ourselves on the absorption opacity. Dust grains of sizes larger
than a few 100 micron will, however, have a substantial scattering albedo
\citep[see discussion on the DSHARP opacity model in][]{dsharp:birnstiel}. How
this affects the results is discussed in Appendix \ref{sec-incl-scattering}.

\section{Effect of scattering albedo}
\label{sec-incl-scattering}
If the dust grains have a radius $a$ comparable to the wavelength of our
observations, the scattering albedo can be quite high. This means that the
absorption optical depth can be substantially lower than unity, even if the full
extinction optical depth (absorption plus scattering) is unity or larger.  The
extinction of $\tau\simeq 0.65$ for ring 1 and $\tau\simeq 0.75$
  for ring 2 found in HD 163296 by \citet{dsharp:isella} from the CO maps could
thus be compatible with the absorption optical depth of $\tau\simeq
  0.44$ for ring 1 and $\tau\simeq 0.33$ for ring 2 we derived in our Gaussian
fitting procedure of the thermal dust emission (see Table
\ref{tab-gauss-params}).

In fact, staying with the case of HD 163296, if we would assume that the albedo
is zero, i.e.\ that the measured extinction optical depth from the CO maps
equals the absorption optical depth, then we would find rather low dust
temperatures at the location of the rings, which may be hard to explain
theoretically.
If, however, part of the extinction is due to scattering, then it is easier
to remain consistent with the dust temperature estimated from the flaring
angle recipe.
\todo{Check if the 0.65 and 0.75 optical depths found by
Andrea are still inconsistent with the temperature.}

However, when scattering is included, the radiative transfer becomes more
complex than a simple use of a factor $1-e^{-\tau_\nu}$. In
\citet{dsharp:birnstiel} we describe an approximate solution to
this problem for a thin slab model. In principle one would have to replace, in
the above sections\removed{} all
instances of $1-e^{-\tau_\nu}$ with the more detailed radiative transfer
\afterintrev{model of \citet{dsharp:birnstiel}}.

\section{Steady-state dust distribution in a ringlike trap}
\label{sec-steady-state-analytic-trap-model}
\subsection{Analytic approximate solution of dust trapping}
\label{sec-analytic-model-of-trapping}
Let us consider a narrow gas ring around the star at radius $r_0$ with a
midplane pressure given by
\begin{equation}\label{eq-gaussian-pressure-bump-repeat}
p(r) = p_0 \exp\left(-\frac{(r-r_0)^2}{2w^2}\right)
\end{equation}
where $w\ll r_0$ is the parameter setting the width of this gaussian gas ring.
We assume that the gas is turbulent with turbulent diffusion coefficient
$D$. Dust grains get trapped in this ring, and the dust will acquire a
radial density profile that is in equilibrium between the radial dust drift
pointing toward the peak of the gas pressure and radial turbulent diffusion
pointing away from that position. The radial dust drift velocity is
\citep[see e.g.][]{2010A&A...513A..79B}:
\begin{equation}\label{eq-v-radial-drift}
  v_{\mathrm{dr}} = \frac{\mathrm{1}}{1+\mathrm{St}^{2}}v_{\mathrm{gr}} +
  \frac{\mathrm{St}}{1+\mathrm{St}^{2}}
  \left(\frac{d\ln p}{d\ln r}\right)\frac{c_s^2}{\Omega_Kr}
\end{equation}
where $c_s$ is the isothermal sound speed and the Stokes number $\mathrm{St}$ is
defined as
\begin{equation}\label{eq-definition-stokes-number}
\mathrm{St} = \Omega_Kt_{\mathrm{stop}}
\end{equation}
where $t_{\mathrm{stop}}$ is the stopping time of the grains. We assume that the
gas radial velocity is zero: $v_{\mathrm{gr}}=0$, but we will briefly discuss
below how the solution shifts slightly away from the peak of the pressure bump
for $v_{\mathrm{gr}}\neq 0$.

The diffusion
coefficient for the dust is \citep{2007Icar..192..588Y}:
\begin{equation}
D_{\mathrm{d}} = \frac{D}{1+\mathrm{St}^2}
\end{equation}
We take $D$ to be equal to the turbulent viscosity $\nu$ divided by the
Schmidt number $\mathrm{Sc}$, which we usually set to $\mathrm{Sc}=1$.
We use the usual $\alpha$-prescription for the turbulence:
\begin{equation}\label{eq-d-afo-alpha}
D=\frac{\nu}{\mathrm{Sc}}= \alpha_{\mathrm{turb}}\frac{c_s^2}{\mathrm{Sc}\,\Omega_K}
\end{equation}
If $D$ is sufficiently small, the dust will get concentrated into a ring with
width $w_d$ that is substantially smaller than the width of the gas ring
$w$. In the following, we will ignore any terms
arising from the curvature of the coordinates. The steady-state radial
dift-mixing equation for the dust then becomes, in its approximate form:
\begin{equation}
  \frac{d}{dr}
  \left(\Sigma_{\mathrm{d}}v_{\mathrm{dr}}-D_{\mathrm{d}}\frac{d\Sigma_{\mathrm{d}}}{dr}\right) = 0
\end{equation}
Integrating this equation once, with integration constant zero (which
amounts to a zero net radial flux), yields
\begin{equation}\label{eq-drift-mix-equil-eq}
  \Sigma_{\mathrm{d}}v_{\mathrm{dr}} = D_{\mathrm{d}}\frac{d\Sigma_{\mathrm{d}}}{dr}
\end{equation}
From Eqs.(\ref{eq-v-radial-drift},\ref{eq-gaussian-pressure-bump-repeat}) we can express
$v_{\mathrm{dr}}$ as
\begin{equation}
v_{\mathrm{dr}} = -\left(\frac{c_s^2}{w^2\Omega_K(\mathrm{St}+\mathrm{St}^{-1})}\right)(r-r_0)
\end{equation}
With this expression we can solve Eq.~(\ref{eq-drift-mix-equil-eq}) for
$\Sigma_{\mathrm{d}}$, leading to the following simple analytic solution to the dust
trapping problem:
\begin{equation}\label{eq-analytic-sol-radial-trapping}
\Sigma_{\mathrm{d}}(r) = \Sigma_{\mathrm{d0}} \exp\left(-\frac{(r-r_0)^2}{2w_{\mathrm{d}}^2}\right)
\end{equation}
with 
\begin{equation}\label{eq-app-wd-in-w}
  w_{\mathrm{d}} = w\, \sqrt{\frac{\Omega_KD_{\mathrm{d}}(\mathrm{St}+\mathrm{St}^{-1})}{c_s^2}}
  = w\,\sqrt{\frac{\alpha_{\mathrm{turb}}}{\mathrm{Sc}\,\mathrm{St}}}
\end{equation}
As a side remark, we note that if $v_{\mathrm{gr}}$ is non-zero and
inward-pointing, this solution shifts inward. We then replace $(r-r_0)$
in Eq.~(\ref{eq-analytic-sol-radial-trapping}) by $(r-r_0-\delta r)$, with
\begin{equation}
  \delta r=\frac{w_d^2v_{\mathrm{gr}}}{D_{\mathrm{d}}(1+\mathrm{St}^2)}
  \simeq \left(\frac{w}{h_p}\right)^2\left(\frac{v_{\mathrm{gr}}}{v_K}\right)
  \frac{1}{\mathrm{St}}\,r_0
\end{equation}
One can see that this shift is independent of the width of the dust ring set by
the turbulence. Note that in the above shift it is assumed that
$v_{\mathrm{gr}}$ is constant across the pressure bump, which breaks mass
conservation for the gas. The above treatment of $v_{\mathrm{gr}}\neq 0$ is
therefore only a rough approximation. We will from here onward return to our
assumption that $v_{\mathrm{gr}}=0$.
  
The normalization constant $\Sigma_{\mathrm{d0}}$ in
Eq.~(\ref{eq-analytic-sol-radial-trapping}) can be approximately expressed in
terms of the total dust mass trapped in the pressure bump:
\begin{equation}
  M_d = 2\pi \int_0^\infty \Sigma_{\mathrm{d}}(r) rdr \simeq
  2\pi r_0 \int_0^\infty \Sigma_{\mathrm{d}}(r)dr
\end{equation}
which leads to
\begin{equation}
\Sigma_{d0}\simeq \frac{M_d}{(2\pi)^{3/2}\, r_0\,w_{\mathrm{d}}}
\end{equation}
The approximation is best for narrow dust rings.

Note that this analytic solution is only valid as long as
$\alpha_{\mathrm{turb}}\ll \mathrm{Sc}\,\mathrm{St}$, or in other words as long
as $w_{\mathrm{d}}$ is substantially smaller than $w$.  This solution is, in
fact, the radial version of the vertical settling-mixing equilibrium solutions
of \citet{1995Icar..114..237D}.

Unfortunately, the condition that $\alpha_{\mathrm{turb}}\ll
\mathrm{Sc}\,\mathrm{St}$ (and equivalently $w_d\ll w$) is easily broken for
small grains and/or non-weak turbulence. In that case our assumption of
a constant $\mathrm{St}$ becomes invalid. Dust will be turbulently mixed
to distances $|r-r_0|\gtrsim w$, where the Stokes number of the grains
increases due to the decreasing gas density. This invalidates
the simple Gaussian solution, at least in principle. 

Given the similarity between the radial dust trapping problem and the vertical
settling problem, one can show that the radial version of the solution of
\citet{2009A&A...496..597F} reads:
\begin{equation}\label{eq-analytic-sol-radial-trapping-better}
  \Sigma_{\mathrm{d}}(r) = \Sigma_{\mathrm{d0}} \exp\Bigg[
    -\frac{\mathrm{Sc}\,\mathrm{St}_0}{\alpha_{\mathrm{turb}}}
     \left(\exp\left(\frac{\Delta r^2}{2w^2}\right)-1\right)
    -\frac{\Delta r^2}{2w^2}\Bigg]
\end{equation}
where we defined $\Delta r$ as
\begin{equation}
\Delta r \equiv (r-r_0)
\end{equation}
and $\mathrm{St}_0$ is the value of the Stokes number at the peak of the
pressure bump.  The solution Eq.~(\ref{eq-analytic-sol-radial-trapping-better})
is valid for any value of $\alpha_{\mathrm{turb}}/\mathrm{Sc}\,\mathrm{St}_0$,
as long as $\alpha_{\mathrm{turb}}$ and $\mathrm{Sc}$ remain constant along the
radial width of the dust trap, the grains remain in the Epstein regime, and
$w\ll r_0$, to prevent geometric terms from the cilindrical coordinates
from dominating.  One can easily verify that
Eq.~(\ref{eq-analytic-sol-radial-trapping-better}) reproduces the simpler
Gaussian solution Eq.~(\ref{eq-analytic-sol-radial-trapping}) for $\Delta r\ll
w$. One can also verify that for $\alpha_{\mathrm{turb}}\gg
\mathrm{Sc}\,\mathrm{St}_0$ the shape of $\Sigma_{\mathrm{d}}(r)$ follows the
shape of the gas pressure profile $p(r)$
(Eq.~\ref{eq-gaussian-pressure-bump-repeat}).

Although this solution is more complete than the simple Gaussian solution,
it turns out that the differences are only in the very wings of the profile.
It will be very hard, if not impossible, for ALMA to distinguish.

For that reason we will in this paper stay with the simpler solution. To
allow the simpler solution to also remain reasonably valid for high
turbulent strength, we will replace Eq.~(\ref{eq-app-wd-in-w}) with
$w_{\mathrm{d}} = w\,\left(1+\psi^{-2}\right)^{-1/2}$, where $\psi$ given by
$\psi = \sqrt{\alpha_{\mathrm{turb}}/\mathrm{Sc}\,\mathrm{St}}$, which turns
out to be a very good approximation.

\section{Stability of gas ring}
\label{sec-gas-ring-stability}
\newstuff{
We have assumed a simple model of a pressure bump: a Gaussian radial pressure
profile given by Eq.~(\ref{eq-gaussian-pressure-bump}). However, it is
known that if the radial pressure gradient is too steep, a Rossby wave
instability can occur \citep{2000ApJ...533.1023L}, which will destroy the axial
symmetry of the ring. The stability of Gaussian gas rings in a potential well
has been studied extensively by \citet{2016ApJ...823...84O}. From their Figure 6
it can be inferred that for the ring to remain stable, it cannot be much
narrower than its vertical extent.
}

\newstuff{
  Let us quantify this using the Solberg-Hoiland stability criterion. Define
  SH as
\begin{equation}
\mathrm{SH} = \kappa^2 + N^2
\end{equation}
If $\mathrm{SH}>0$ then the disk is stable. If $\mathrm{SH}<0$ then the disk is
unstable. We follow \citet{2000ApJ...533.1023L}, their Eq.(22), though
with midplane density and pressure. The $\kappa$ is given by the derivative of
the specific angular momentum in the following way:
\begin{equation}
\kappa^2 = \frac{1}{r^3}\frac{dl^2}{dr}
\end{equation}
where $l=v_\phi r$ is the angular momentum of the gas. Due to the pressure
gradient, this is not exactly the Keplerian angular momentum, but:
\begin{equation}
l^2 = l_K^2 + c_s^2r^2\left(\frac{d\ln p}{d\ln r}\right)
\end{equation}
where $l_K=\Omega_Kr^2$.
The Brunt-Vaisala frequency is given by:
\begin{equation}
  N^2    = \frac{1}{\rho}\frac{dp}{dr}\left(\frac{1}{\rho}\frac{d\rho}{dr}-
  \frac{1}{\gamma p}\frac{dp}{dr}\right)
\end{equation}
where $\gamma$ is the adiabatic index. Let us, for the sake of simplicity,
assume that the dimensionless scale height of the disk, $h_p/r$, is constant
with $r$, which implies that $c_s^2\propto T\propto 1/r$. The pressure
profile is given by Eq.~(\ref{eq-gaussian-pressure-bump}). With some algebra
we find:
\begin{eqnarray}
  \kappa^2 &=& \Omega_k^2\left[1-\left(\frac{h_p}{w}\right)^2\left(3-2\frac{r_0}{r}\right)\right]\\
  N^2 &=& \frac{c_s^2}{r^2}\frac{r(r-r_0)}{w^2}\left[
    \frac{r(r-r_0)}{w^2}\left(1-\frac{1}{\gamma}\right)-1\right]
\end{eqnarray}
This leads us to
\begin{equation}
  \frac{\mathrm{SH}}{\Omega_K^2} = 1-\left(\frac{h_p}{w}\right)^2
  \left\{4-3\frac{r_0}{r}-\frac{(r-r_0)^2}{w^2}\left(1-\frac{1}{\gamma}\right)\right\}
\end{equation}
Close to $r_0$ the first two terms between the $\{\}$ brackets are roughly 1.
For $\gamma=7/5$ we get $1-1/\gamma=2/7$.
}

\newstuff{
We see that if $h_p\lesssim w$, then the Gaussian pressure bump is stable
($\mathrm{SH}>0$). However, for $h_p\gtrsim w$ we find $\mathrm{SH}<0$, and the ring becomes
unstable.
}

\todo{Check the above algebra.}

\section{An effective 1-D kernel consistent with 2-D beam convolution}
\label{sec-effective-kernel}
The convolution of the emission from a 1-D axisymmetric disk model is a 2-D
process due to the inclination of the disk and the ellipticity of the
interferometrically synthesized beam. That means that, in order to compare such
a 1-D model to the data, we need to convert it into a 2-D model (or even a 3-D
model if the disk's vertical thickness is non-negligible), and then put it at an
inclination, project it onto the sky, and perform a 2-D convolution with the
elliptic beam. This image can then be compared to the measured image. While
straightforward, this is a computationally costly procedure.

For the limiting case of a geometrically extremely thin layer of thermally
emitting dust it is, however, possible to describe this 2-D convolution
procedure analytically, as long as we focus on radii $r$ much larger than the
beam size. This leads to an ``effective 1-D convolution kernel'' that can be
applied directly to the 1-D model emission and compared directly to the 1-D
radial intensity profiles extracted from the observations. 

The procedure involves a linear average of convolutions along radial rays in the
image plane. Due to the inclination of the disk and the ellipticity of the beam,
each of these convolutions smears out ringlike structures in the disk to a
different degree. Typically the smearing, relative to the radial coordinate $r$
in the disk plane, is more severe along the minor axis of an inclined disk by a
factor $1/\cos i$ compared to the major axis. Likewise it is more severe along
the major axis of the beam by a factor
$\sigma_{\mathrm{maj}}/\sigma_{\mathrm{min}}$ compared to the minor axis, where
$\sigma_{\mathrm{maj}}$ and $\sigma_{\mathrm{min}}$ are the standard deviation
beam widths along the major and minor axis of the beam, respectively.

We first deproject the annulus, thereby stretching
the beam in the direction of the minor axis of the disk. Then we perform a
linear coordinate transformation to make the beam circular again. The annulus
has, by then, become elliptic once more. The task is to calculate the width of
the segment of the annulus in this skewed coordinate system along a given ray. The
relative width of the circularized beam to the width this segment is a measure
of how strongly the beam affects the annulus along this ray.

The averaging will be done in the coordinate $\phi$, which is the azimuthal
coordinate in the plane of the disk. We denote the inclination as $i$, the
position angle of the disk's major axis as $\alpha$, measured east-of-north.
The position angle of the beam is denoted as $\xi$, and is defined in the
same manner as $\alpha$. The azimuthal coordinate $\phi$ is clockwise when
viewed at inclination $i=0$, and $\phi=0$ lies along the minor axis, east
of the center when $\alpha=0$. These definitions are the same as used
by \citet{dsharp:huangspirals}.

We start with an annulus width of $\delta r$ in the plane of the disk, the
annulus being the radial range $[r,r+\delta r]$. After deprojection this
width has changed to
\begin{equation}
\delta r' = \delta r \,\frac{|\cos i|}{\sqrt{\cos^2\phi+\cos^2 i\,\sin^2\phi}}
\end{equation}
This projection also changes the angle of the annulus segment on the sky. If
$\beta=\phi$ is the original angle between the segment and the major axis
of the projected disk, then the new angle $\beta'$ obeys $\tan\beta' = \cos
i\,\tan\beta$. Next we rotate the coordinate system such that the elliptic beam
lies horizontal. The new angle of the annulus segment $\beta''$ is now
$\beta''=\beta'+\xi-\alpha-\pi/2$, measured clockwise from positive x-axis.
The final projection leads to a width:
\begin{equation}\label{eq-delta-r-pp}
  \delta r'' = \delta r' \,\frac{\sigma_{\mathrm{min}}/\sigma_{\mathrm{maj}}}
         {\sqrt{\cos^2\beta''+(\sigma_{\mathrm{min}}/\sigma_{\mathrm{maj}})^2\,\sin^2\beta''}}
\end{equation}
From this we can say that the smearing-out of the annulus segment by the beam
(the ratio by which the beam segments gets wider by the convolution) is
$(\delta r/\delta r'')$ times stronger than if a circular beam with
$\sigma_{\mathrm{min}}\times \sigma_{\mathrm{min}}$ would be applied in the
deprojected disk plane. In the coordinate $r$ the radial beam standard
deviation width along this ray is then
\begin{equation}
\sigma_{\mathrm{ray}}(\phi) = \frac{\delta r}{\delta r''}\,\sigma_{\mathrm{min}}
\end{equation}

The effective 1-D convolution kernel, to be used in conjunction with the
$r$-coordinate in the disk plane, is then:
\begin{equation}
  K_{\mathrm{eff}}(r'-r,i) =
  \frac{1}{2\pi} \int_0^{2\pi}
    K\big(r'-r,i,\sigma_{\mathrm{min}}\delta r/\delta r''(\phi)\big)\, d\phi
\end{equation}
where $K(r'-r,i,\sigma_b)$ is the Gaussian kernel with standard deviation
$\sigma_b$. The 2-D convolution then becomes again a 1-D convolution, but with 
the effective kernel:
\begin{equation}
  I_\nu^{\mathrm{conv}}(r,i) =  \int_0^\infty 
    I_\nu(r',i) K_{\mathrm{eff}}(r'-r,i)\, dr'
\end{equation}

\revised{In most cases this complex effective kernel can be approximated
  fairly well with a Gaussian kernel with average width given by:
\begin{equation}\label{eq-average-beam-width}
\sigma_{\mathrm{av}} = \sqrt{\frac{\sigma_{\mathrm{min}}\sigma_{\mathrm{maj}}}{|\cos i|}}
\end{equation}
Only when the disk has a large inclination and the beam is strongly elliptic will
this approximation fail.
}

\section{Mock ring test}
\label{sec-mock-ring-test}
\newstuff{
Strictly speaking, comparing a model to interferometric data is best done in the
uv-plane. But the high quality of the ALMA data allows also a model comparison
in the image plane. The advantage is that one can select individual features
while ignoring the rest. In this paper we analyze our data close to the spatial
resolution limit. To check the reliability of this, we perform here a simple
test: We set up a single mock ring inspired by ring 1 (B74) of AS 209, with
the width $w_d=3.07\,\mathrm{au}$ from \citet{dsharp:guzman}, add some
reasonable noise, simulate the ALMA visibilities, put these data through the
DSHARP imaging pipeline, and extract the radial profile. We compare this result to a
simple 2-D convolution of the mock ring, as well as to the 1-D convolution with
the effective kernel discussed in Appendix \ref{sec-effective-kernel}.}

\begin{figure}
\centerline{\includegraphics[width=0.5\textwidth]{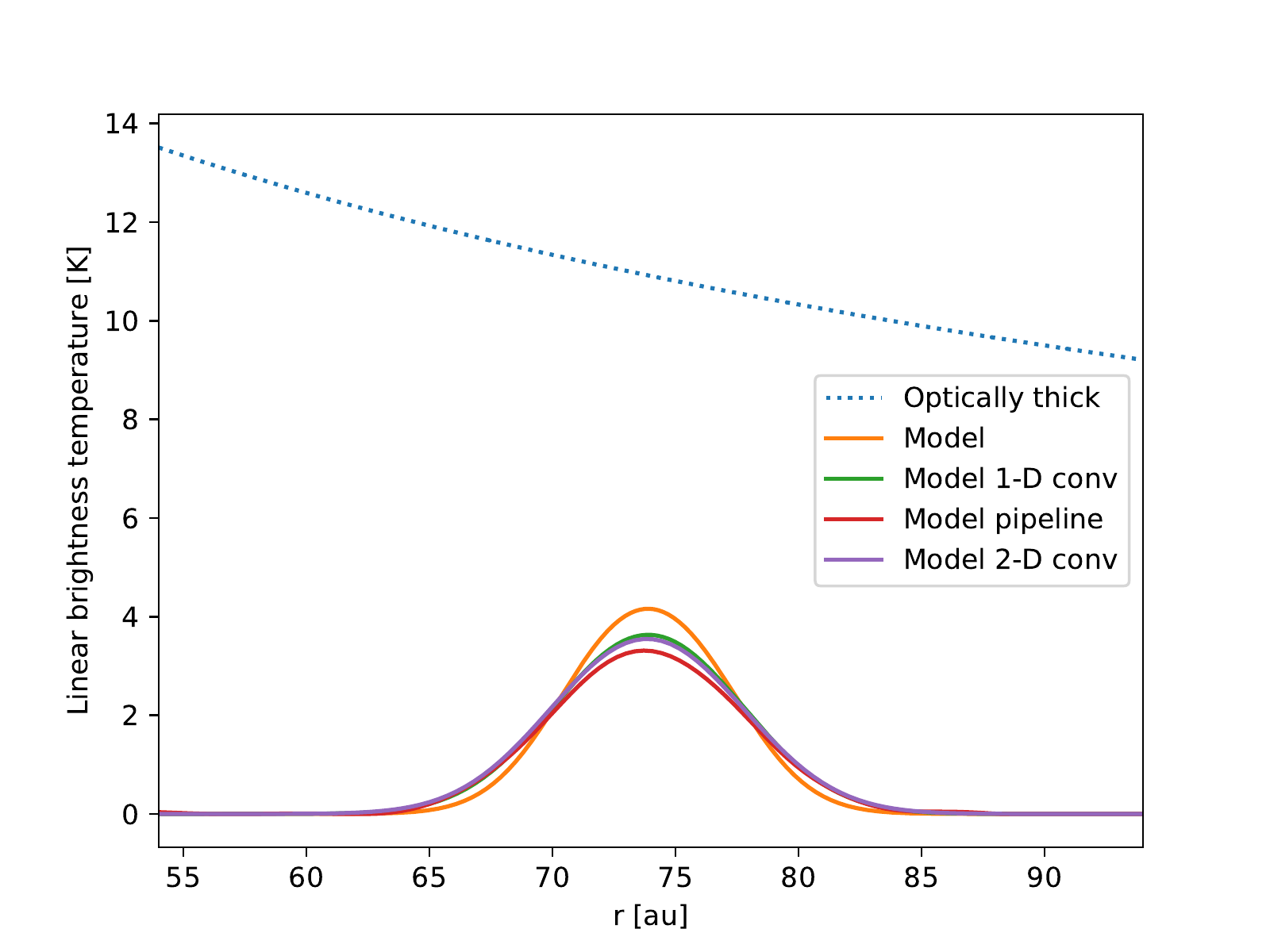}}
\caption{\label{fig-mock-1d-extractions}The 1-d radial profiles extracted from
  the 2-D images in Fig.~\ref{fig-mock-images}, compared to the original mock
  ring.}
\end{figure}

\begin{figure*}
\centerline{\includegraphics[width=0.95\textwidth]{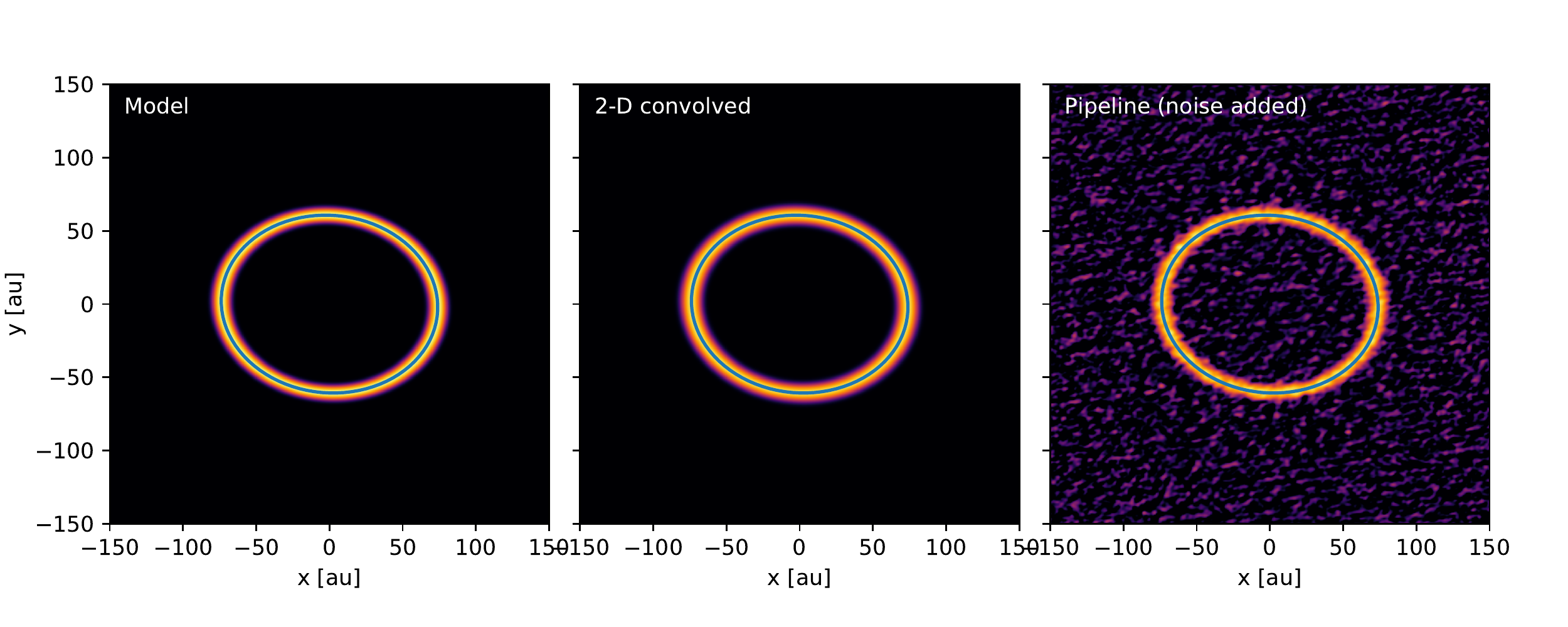}}
\caption{\label{fig-mock-images}The mock ring test. Left: the input mock ring,
  inspired by ring 2 of AS 209, assuming a width of
  $w_d=3.07\,\mathrm{au}$. Inclination and position angle are the same as for AS
  209. Middle: The mock ring convolved with the Gaussian beam appropriate for AS
  209. Right: The mock ring, with noise added, put through the DSHARP imaging
  pipeline.}
\end{figure*}

\newstuff{The mock ring, its 2-D convolved version and the end result of the
  imaging pipeline (after noise was added) are shown in Fig.~\ref{fig-mock-images}. The
  resulting 1-D extractions are shown in Fig.~\ref{fig-mock-1d-extractions}. The
  optical depth effects made the unconvolved mock ring emission a bit wider than
  the underlying dust ring: $\sigma=3.27\,\mathrm{au}$. For the 1-D-convolved
  ring (using the effective kernel) we find
  $\sigma_{\mathrm{conv}}=3.76\,\mathrm{au}$, for the 2-D-convolved ring we find
  $\sigma_{\mathrm{conv}}=3.85\,\mathrm{au}$, and for the full pipeline we find
  $\sigma_{\mathrm{conv}}=3.86\,\mathrm{au}$.}

\newstuff{These results show that in principle there should be no appreciable
  difference between the spreading of the emission by the simulated observation
  and the 2-D and 1-D convolutions.}

\newstuff{The fact that the uv-plane fitting results of \citet{dsharp:guzman}
  for AS 209, \citet{dsharp:isella} for HD 163296, and \citet{dsharp:perez} for
  HD 143006 result in widths that are not exactly the same as in this paper may
  be due to the different fitting criteria used. The fitting in the present
  paper focuses on the shape near the peak of the radial intensity profile,
  while the fitting in the uv plane acts on the the full dataset. Whether this
  fully explains the differences remains unclear.}

\end{document}